\documentclass[useAMS,usenatbib]{mnras}

\usepackage{graphicx}
\usepackage{color}
\usepackage{multirow}
\usepackage{amssymb,amsmath}
\usepackage{cprotect}

\def\aap{A\&A}
\def\apj{ApJ}
\def\araa{ARAA}
\def\apjl{ApJL}
\def\aj{AJ}
\def\pasp{PASP}
\def\pasa{PASA}
\def\mnras{MNRAS}
\def\nat{Nature}
\def\na{New Astron.}

\def\apjs{ApJS}

\def\ps{$\rm km \,s^{-1}\,kpc^{-1}$}

\def\arcdeg{$^\circ$}
\def\kms{\,km\,s$^{-1}$}

\def\H2{$\rm H_2$}

\def\htwo{H{\small II}\,}
\def\delv{$\nabla \cdot v_{xy}$}

\title[Young stars in the Milky Way]{Young stars as tracers of a barred-spiral Milky Way}

\author[A. R. Pettitt et al.]
{Alex R. Pettitt$^{1}$\thanks{E-mail: alex@astro1.sci.hokudai.ac.jp}, Sarah E. Ragan$^2$ and Martin C. Smith$^3$,\\
$^{1}$Department of Physics, Faculty of Science, Hokkaido University, Sapporo 060-0810, Japan\\
$^{2}$School of Physics and Astronomy, Cardiff University, Queen’s Buildings, The Parade, Cardiff, CF24 3AA, UK\\
$^{3}$Key Laboratory for Research in Galaxies and Cosmology, Shanghai Astronomical Observatory, Chinese Academy of Sciences,\\ 80 Nandan Road, Shanghai 200030, People's Republic of China
}

\begin{document}

\date{\today}

\pagerange{\pageref{firstpage}--\pageref{lastpage}} \pubyear{201X}

\maketitle
\label{firstpage}
 
\begin{abstract}
{Identifying the structure of our Galaxy has always been fraught with difficulties, and while modern surveys continue to make progress building a map of the Milky Way, there is still much to understand. The arm and bar features are important drivers in shaping the interstellar medium, but their exact nature and influence still require attention. We present results of smoothed particle hydrodynamic simulations of gas in the Milky Way including star formation, stellar feedback, and ISM cooling, when exposed to different arm and bar features, with the aim of better understanding how well newly formed stars trace out the underlying structure of the Galaxy. The bar is given a faster pattern speed than the arms, resulting in a complex, time-dependent morphology and star formation. Inter-arm branches and spurs are easily influenced by the bar, especially in the two-armed spiral models where there is a wide region of resonance overlap in the disc. As the bar over-takes the spiral arms it induces small boosts in star formation and enhances spiral features, which occur at regularly spaced beat-like intervals. The locations of star formation events are similar to those seen in observational data, and do not show a perfect 1:1 correspondence with the underlying spiral potential, though arm tangencies are generally well traced by young stars. Stellar velocity fields from the newly formed stars are compared to data from Gaia DR2, showing that the spiral and bar features can reproduce many of the non-axisymmetric features seen in the data.  A simple analytical model is used to show many of these feature are a natural response of gas to rigidly rotating spiral and bar potentials.
}\end{abstract}


\section{Introduction}

Our home galaxy, the Milky Way, is of keen interest for studies of both large scale galactic dynamics and smaller scale properties of the interstellar medium (ISM). A tremendous quantity of high quality observational data is being gathered in hopes of better understanding the structure of the Milky Way's stellar and gaseous media \citep{2015MNRAS.449.2604D,2017PASJ...69...78U,2017A&A...601A.124S,2018A&A...616A...1G}. This has brought with it remarkable new insights (e.g. \citealt{2018A&A...616A..11G,2018Natur.561..360A}), further reinforcing a need to fully understand the underlying non-axisymmetric features of the Galaxy: the spiral arms and inner bar. See recent reviews of \citet{2016ARA&A..54..529B} and \citet{2018RAA....18..146X} for contemporary discussions of our current understanding of Milky Way structure. 

Our basic picture of the Milky Way's non-axisymmetric structure is that there are 2--4 primary spiral arms (though an abundance of inter-arm structure persists) with some inner bar feature (be it long, thin, short, boxy etc.). This interpretation comes from a combination of both observational evidence \citep{1976A&A....49...57G,2000A&A...358L..13D,2009PASP..121..213C} and inferred by constraining theoretical models \citep{1999A&A...345..787F,2005AJ....130..576Q,2010PASJ...62.1413B}. Additionally, these non-axisymmetric features of the Milky Way are believed to rotate at different pattern speeds. The general consensus is that the bar rotates faster than the spiral arms \citep{2003MNRAS.340..949B,2011MSAIS..18..185G, 2014arXiv1406.4150P,2015MNRAS.449.2336J,2016ApJ...824...13L,2017Ap&SS.362...79V}, though how much faster is still the subject of theoretical and observational efforts \citep{2014A&A...563A..60A,2015MNRAS.448..713P,2015MNRAS.454.1818S}. Summarising some of the above works: the Milky Way's spiral arms are believed to rotate at speeds between $10$\ps{} to $30$\ps{}, while the bar rotates somewhere between the spiral arm pattern speed and 70\ps{}, with contemporary measurements putting the speed at around 40\ps{}  \citep{2017MNRAS.465.1621P,2019arXiv190511404B}. Observations of extragalactic systems also indicate bars are significantly faster rotators than arms \citep{2011MSAIS..18...23C,2016MNRAS.459.4057G,2017ApJ...835..279F,2019MNRAS.482.5362F}.

The mis-match of spiral and bar pattern speeds can cause a variety of interesting features in the disc. For example, \citet{2006MNRAS.368..623M} look at mismatching pattern speeds (between a 2 and 4 armed potential) and see they act together to strongly and non-linearly increases the velocity dispersion, causing a diffusion in phase-space. This was followed up by specifically looking into bar-spiral overlapping resonant features, where the authors show the overlap can induce radial migration on Gyr timescales \citep{2010ApJ...722..112M,2011A&A...527A.147M}. However, more recent work by \citet{2016MNRAS.461.3835M} indicates this resonant overlap may not induce such large perturbations in the stellar dynamics. While the effect of bar-spiral coupling on stellar kinematics has been the subject to several past investigations, little attention has been paid to the effect on the ISM and resulting star formation.

The multi-phase ISM offers some of the best constraints on Galactic structure, either through construction of top-down maps (e.g. \citealt{1958MNRAS.118..379O,2006Sci...312.1773L,2009A&A...499..473H}) or longitude-velocity diagrams (e.g. \citealt{2001ApJ...547..792D,2006ApJS..163..145J,2017PASJ...69...78U}). Dense molecular gas emission, and giant molecular clouds themselves clearly trace out some morphological features of the Milky Way \citep{2009ApJ...699.1153R,2014ApJS..212....2G,2015ARA&A..53..583H,2016MNRAS.456.2885R}, though the data hardly seems to point to a clear underlying grand design. Such tracers of the ISM have been used to constrain theoretical models for many decades \citep{1986A&A...157..148M,1999MNRAS.304..512E,2004ApJ...615..758G,2015MNRAS.446.4186S}. The properties and evolution of the ISM in galaxies is of prime importance for the role it plays in star formation. The evolution of the star forming ISM is complex, with shocks induced by spiral arm passages, cloud-cloud collisions, and stellar feedback playing key roles. Maps of star forming complexes in the Milky Way (e.g. H$\alpha$ emission from \htwo regions and dust emission from embedded star formation) show several arm-like features \citep{2003A&A...397..133R,2014A&A...569A.125H}. It is expected that the ISM and locations of star formation can divulge key insights in validating spiral arm theories, with classical density waves driving offsets from the potential minimum due to shocks in the gas \citep{1964ApJ...140..646L,1969ApJ...158..123R}. While this is seen in some extra-Galactic systems (e.g. \citealt{2017MNRAS.465..460E}), evidence for offsets, gradients in stellar populations or an increased star formation efficiency associated with spiral arms is not forthcoming \citep{2010ApJ...725..534F,2012MNRAS.426..701M,2015MNRAS.452..289E,2018MNRAS.479.2361R}. While some attempts at modelling how the complex bar/spiral morphology of the Milky Way will impact the star forming ISM \citep{2004A&A...421..863M,2006MNRAS.371.1663D,2018A&A...609A..60K}, none have focussed on the role kinematically independent arms and bars may play.

The goal of this paper is to investigate the locations and dynamics of the young stellar population of the Galaxy, specifically looking into how bar and spiral patterns of differing pattern speeds impact such features. For instance, how reliable is the young stellar population as a tracer of Galactic structure in light of such perturbations? We do so by performing simulations of the ISM embedded in analytic background potentials. This paper is organised as follows. Section \ref{sec:numerics} describes the simulation methodology. Section \ref{sec:results} presents our results, and Section \ref{sec:discussion} a discussion in light of modern observational efforts. In Section \ref{sec:model} a comparison is made between a simple analytic model and the simulations results, and we conclude in Section \ref{sec:conc}.

\section{Numerical Simulations}
\label{sec:numerics}
Simulations are performed using the $N$-body, smoothed particle hydrodynamics (SPH) code \textsc{gasoline2} \citep{2004NewA....9..137W,2017MNRAS.471.2357W}. Gravity is solved using a binary tree, and the system integrated using a kick-drift-kick leapfrog. We use 64 neighbours and the standard cubic spline kernel for all simulations presented here. Adaptive softening lengths were used for the gas, and fixed gravitational softening lengths (50\,pc) for the stars. The simulations include stellar feedback, star formation and ISM cooling process, identical to those used in \citet{2017MNRAS.468.4189P}, i.e. using the blastwave approach and tabulated cooling tables \citep{2006MNRAS.373.1074S,2010MNRAS.407.1581S}.

We use the same potentials as in \citet{2014arXiv1406.4150P} to model the Milky Way system, but using a slightly different gas distribution. We employ a more standard exponential surface density disc, rather than the one exactly tailored to the observed Milky Way (lacking the sharp decrease near the galactic centre). The disc has a mass of $6\times 10^{9}M_\odot$ with radial and vertical scale-lengths of 7kpc and 0.4kpc respectively. We perform seven different calculations with an initial gas resolution of 4 million particles (a mass resolution of 1500M$_\odot$ per gas particle). While this resolution is clearly insufficient to resolve individual star forming events, it still allows the capture of global trends seen on Galactic-scales. 

\begin{figure}
\includegraphics[trim = 10mm 0mm -10mm 0mm,width=90mm]{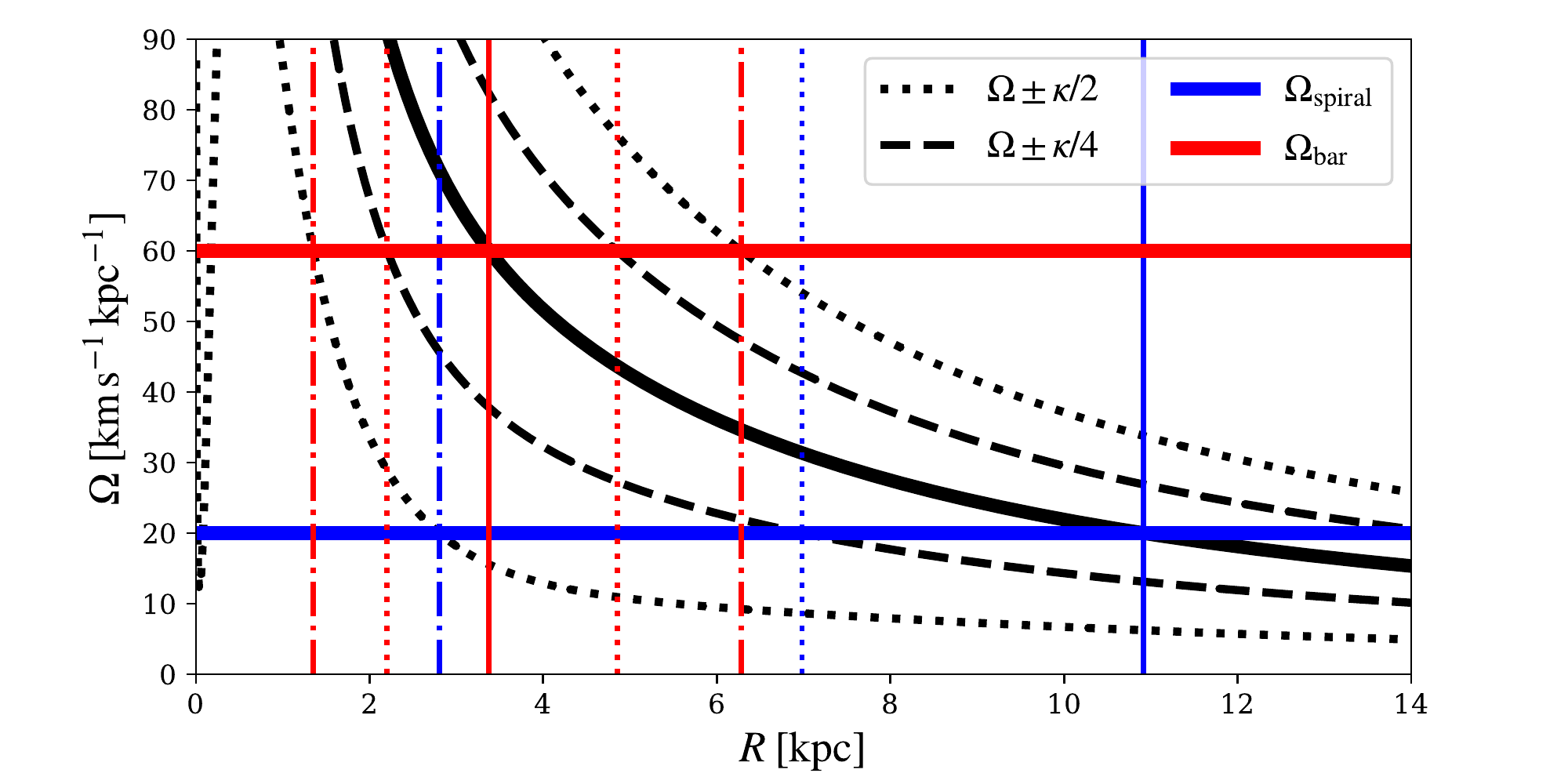}
 \caption{Rotation frequencies for the adopted model. Black lines indicate the material rotation frequencies; solid is $\Omega(R)$, dashed are $\Omega\pm\kappa/4$ and dotted $\Omega\pm\kappa/2$. The blue and red horizontal lines indicate the pattern speed of the arms and bar, with vertical lines showing radii of corresponding resonances (2:1 and 4:1).}
\label{resfig}
\end{figure}

All calculations are subject to an axisymmetric component disc+bulge+halo, identical to that used in \citet{2014arXiv1406.4150P}. We perform a number of different simulations including different non-axisymmetric components. Simulations denoted Br include a bar potential of \citet{2001PASJ...53.1163W}, rotating at a speed of 60\,\ps{} with a scale length of $\sqrt{2}$\,kpc. The model BrStr is a slower and stronger bar, having a pattern speed of $50$\,\ps{} and a strength increased by a factor of $\sqrt{2}$. Bar potentials such as this are known to generate arm features from their co-rotation radius, which behave as kinematic density waves \citep{2015MNRAS.451.3437S}. Simulations denoted Sp include the spiral arms from \citet{2002ApJS..142..261C}, with a pitch angle of 15\arcdeg{} and pattern speed of 20\,\ps{}. Sp2 and Sp4 imply a 2 and 4 armed spiral component, with BrSp models having both bar and spiral features active (with different pattern speeds). All non-axisymmetric components are assumed to rotate as rigid features. While there is increasing evidence that a more dynamic and transient spiral structure exists in disc galaxies from both simulations  \citep{2015MNRAS.449.3911P,2015MNRAS.453.1867G} and observations \citep{2009ApJ...702..277M,2011ApJ...735..101F,2015ApJ...810....9C}, we use fixed potentials for their simplicity and more canonical acceptance in the literature. We note that the nature of spiral arms is still a topic of uncertainty (see \citealt{2014PASA...31...35D} for a review), and different theories have repercussions for disc kinematics \citep{2015MNRAS.453.1867G,2019MNRAS.484.3154S}, radial migration \citep{2002MNRAS.336..785S} and giant molecular cloud properties \citep{2012MNRAS.425.2157D,2017MNRAS.464..246B}.

We use the smooth start activation of \citet{2000AJ....119..800D} for the spiral and bar potentials, where the bar and arms are smoothly activated over 100\,Myr and 300\,Myr respectively. No old stellar population is dynamically modelled, as they are represented by the rigid background potentials. Seven simulations are performed: D, Br, BrStr, Sp2, Sp4, BrSp2, BrSp4, where D has only the axisymmetric potential. Figure\;\ref{resfig} shows the locations of the various resonances associated with the spiral (20\ps{}) and bar (60\ps{}) perturbations for our choice of rotation curve. These include the co-rotation resonance (CR), inner Lindblad (ILR) and outer Lindblad resonance (OLR). The former is simply where the rotation of the rigid spiral/bar pattern matches that of the differentially rotating disc: $\Omega_{p}=\Omega_{\rm gas}(R_{\rm CR})$. The I/OLR are where the disc material undergoes one radial oscillation in passing between symmetric features in the spiral/bar potential. For a bar and two-armed spiral these 2:1 resonances occurs when $\kappa/[\Omega_{p}-\Omega_{\rm gas}(R_{\rm LR})]=\pm 2/1$, and for a four-armed spiral (4:1) when $\kappa/[\Omega_{p}-\Omega_{\rm gas}(R_{\rm LR})]=\pm 4/1$. The 4:1 resonance for 2-fold symmetric structures is often called the ultra-harmonic resonance. Such radii play important roles in gas and stellar kinematics. They can act to reflect and dissipate density waves, act as drivers of angular momentum transfer and disc heating, and can lead to the formation of secondary arm branches and other non-linear features \citep{1973ApJ...183..819S,1978ApJ...222..850G,1986A&A...155...11C,1989ApJ...343..602E}. It is clear there are locations where resonances are very close to each other (e.g. the spiral inner 4:1 and bar outer 2:1), and these overlapping resonances have been shown to drive long-term (order of Gyr) migration and non-linear heating of stellar discs \citep{2010ApJ...722..112M}.

For real external galaxies there is surprisingly little data for the radial dependence of the pattern speeds (i.e. tracing pattern speeds of bars and arms in a single galaxy). Measurements tend to show that pattern speeds of features decrease with radius \citep{2006MNRAS.366L..17M}, and that bars and spiral may be dynamically distinct from the arms \citep{2016ApJ...826....2S}.  $N$-body simulations, however, allow for more definite measurements. Bars tend to be rotating at some rapid rotation speed that is constant with radius, which then transitions into lower pattern speed at the arms \citep{2011MNRAS.417..762Q,2013MNRAS.432.2878R}. The transition is fairly smooth and is often characterised by a breaking and reconnecting between the arms and bar due to the pattern speed mis-match at the transition region \citep{1988MNRAS.231P..25S,2012MNRAS.426..167G,2015MNRAS.454.2954B}. The fact we see arms extending from bars in many galaxies may point to systems where bars are the key dynamical driver, or that we see epochs where arms and bars have re-connected. Indeed, there are also systems where arms and bars show no clear-cut connection, usually in multi-armed, Milky Way type galaxies (e.g. UGC 12158, NGC 1232, M61). We note that spirals in $N$-body simulations also often have a radially decaying pattern speed, with no strong consensus as to whether grand design, single mode, constant pattern speed arms are maintainable in simulations \citep{2014PASA...31...35D}.

Our choice of bar and bulge models means stars/gas are not very well-aligned along the bar major axis itself. However; we are specifically looking at the mid/outer disc response of the gas and stars, rather than the orbits within the bar itself. Other models exist in the literature that can form thinner bars (e.g. \citealt{2016ApJ...824...13L}) but they also induce very small forcing/torque time-steps on particles at the bar end, raising an unnecessary computational hinderance for our study.

\section{Results}
\label{sec:results}

\subsection{General morphology}

\begin{figure*}
\includegraphics[trim = 20mm 15mm 10mm 0mm,width=85mm]{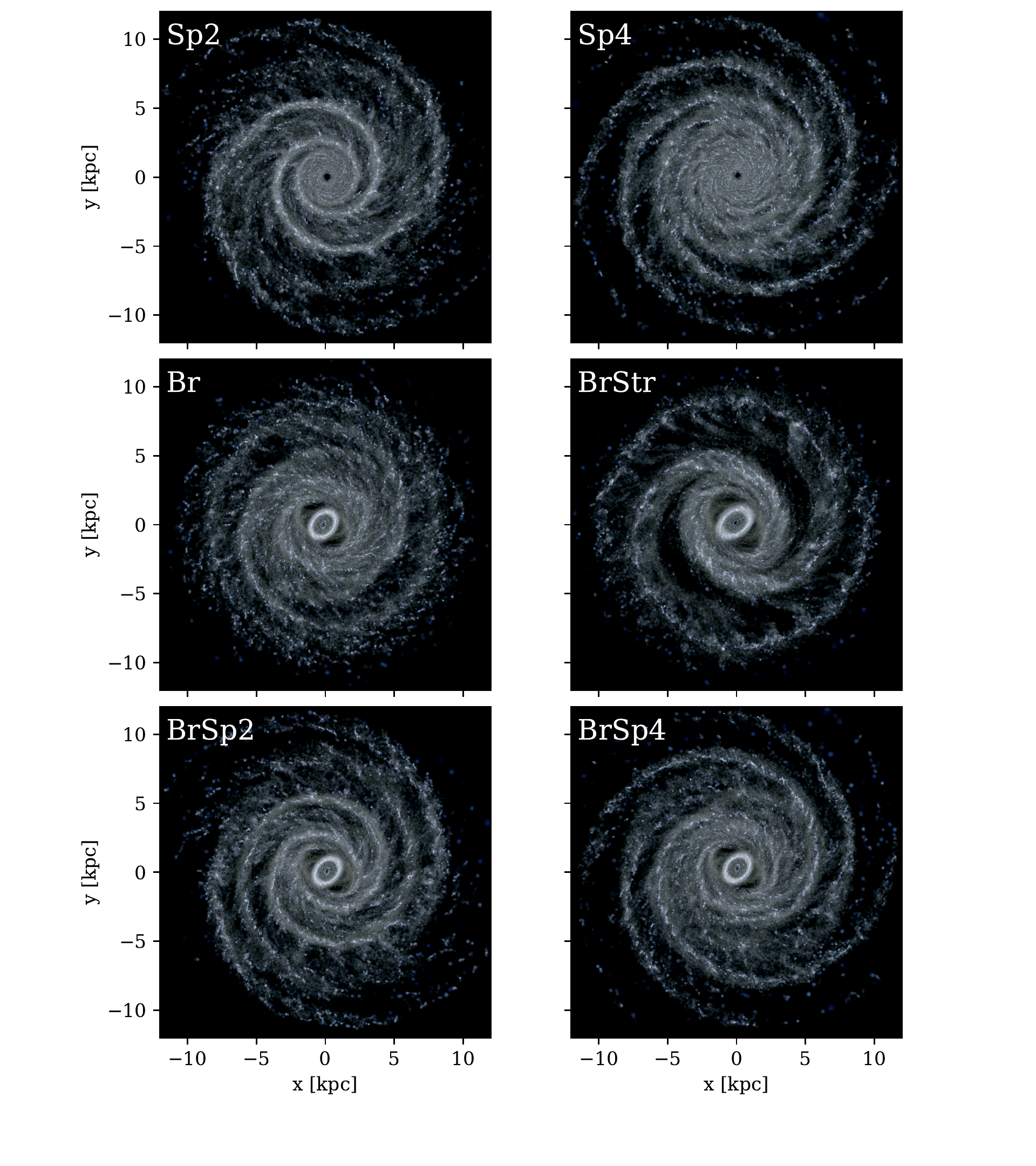}
\includegraphics[trim = 10mm 15mm 20mm 0mm,width=85mm]{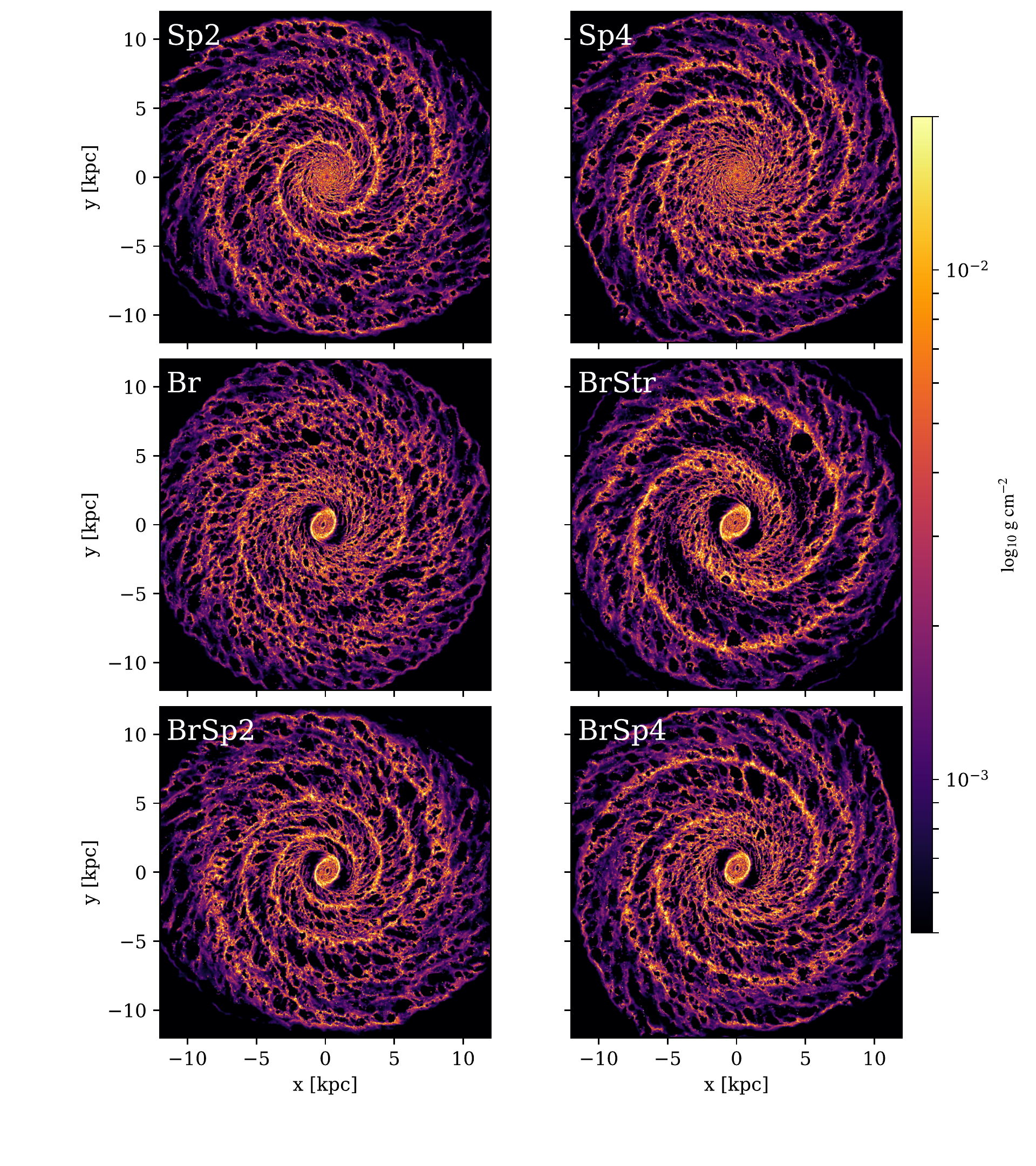}
 \caption{Top-down gas column densities of the six non-axisymmetric models (right) and mock stellar images of the young stellar population (left) after 300\,Myr of evolution. The disc with no bar/spiral perturbation is not shown, as the disc structure is axisymmetric on large scales. Models denoted Br and Sp$N$ include bars and spiral arms, where $N$ is the arm number, and BrStr the slower bar.}
\label{ImsAll_G}
\end{figure*}

The right 6 panels of Figure\;\ref{ImsAll_G} show the top-down gas column density of all models with bar or spiral potentials. Gas begins to uniformly cool and collapse shortly after initialisation, with new star particles forming after only a few 10s of Myr, which then deposit feedback energy back into the ISM, excavating cavities in the gas. The spirals and bars begin to grow as the simulation progresses, with both having reached maximum strength by 300\,Myr. These rotate as rigid features, so gas flows through the perturbations at a rate that changes with galactic radius\footnote{Movies showing the time evolution of the simulations can be found at: \url{https://zenodo.org/record/3239994}.}. 

The two spiral models clearly show dense gas tracing the spiral arm features, with the 2-armed model penetrating deeper into the centre than the 4-armed case due to the contraction of the inner Lindblad resonance caused by different azimuthal symmetry between the two. The weak bar only lightly perturbs the gas disc in the mid/outer regions, whereas the stronger bar clearly has a greater impact. Both of these bars create spiral arm-like features in the gas disc, rotating with the bar potential. The capability of bars to independently drive spiral arm features has been identified in numerous pervious works \citep{1994PASJ...46..165W,1994ASPC...66...29L}. Such spiral features are not necessarily log-spirals, with the major arms existing approximately between CR and the OLR, wrapping to a pitch of 0\arcdeg{} at the latter (agreeing with both Br and BrStr maps) and forming additional arms or rings near the ILR \citep{2012MNRAS.421.1089P}. Such bar-driven spirals are unique to the gas, with dissipation-free stars instead exhibiting orbits either parallel or perpendicular with the bar \citep{2014PASA...31...35D}. The stellar maps of Fig\;\ref{ImsAll_G} exhibit spiral-like features for Br and BrStr because these are only maps of young stars (with even the oldest stars only having experienced a single Galactic rotation), whose orbits are still seeded by that of the progenitor gas particles.

The mixed models with bars and spiral arms show a more complex structure, especially in the BrSp2 model. Here the bar-driven arms and spiral model overlap over a range of radii, creating additional arm features due to their mis-matching pattern speed. There is a periodic matching of bar-driven arms and spiral-driven arms between the bar co-rotation and OLR. BrSp4 is less complex, owing to the 4-armed feature being confined to the outer disc where the bar has little influence. However, the bar in this model does seem to reinforce two of the four arms in particular, naturally creating a stronger arm pair extending from the bar ends and a weaker pair does does not. Similar results were seen in simulations by \citet{2003MNRAS.340..949B}, who adopted the same pattern speed ratio of 3:1 and see similar features in the gas. The features of the four armed mixed model, BrSp4, is akin to what has been inferred for the Milky Way's spiral arms in the 4-armed paradigm, with two out of the four arms appearing stronger \citep{2000A&A...358L..13D,2009PASP..121..213C}.

\begin{figure}
\includegraphics[trim = 0mm 10mm 0mm 0mm,width=80mm]{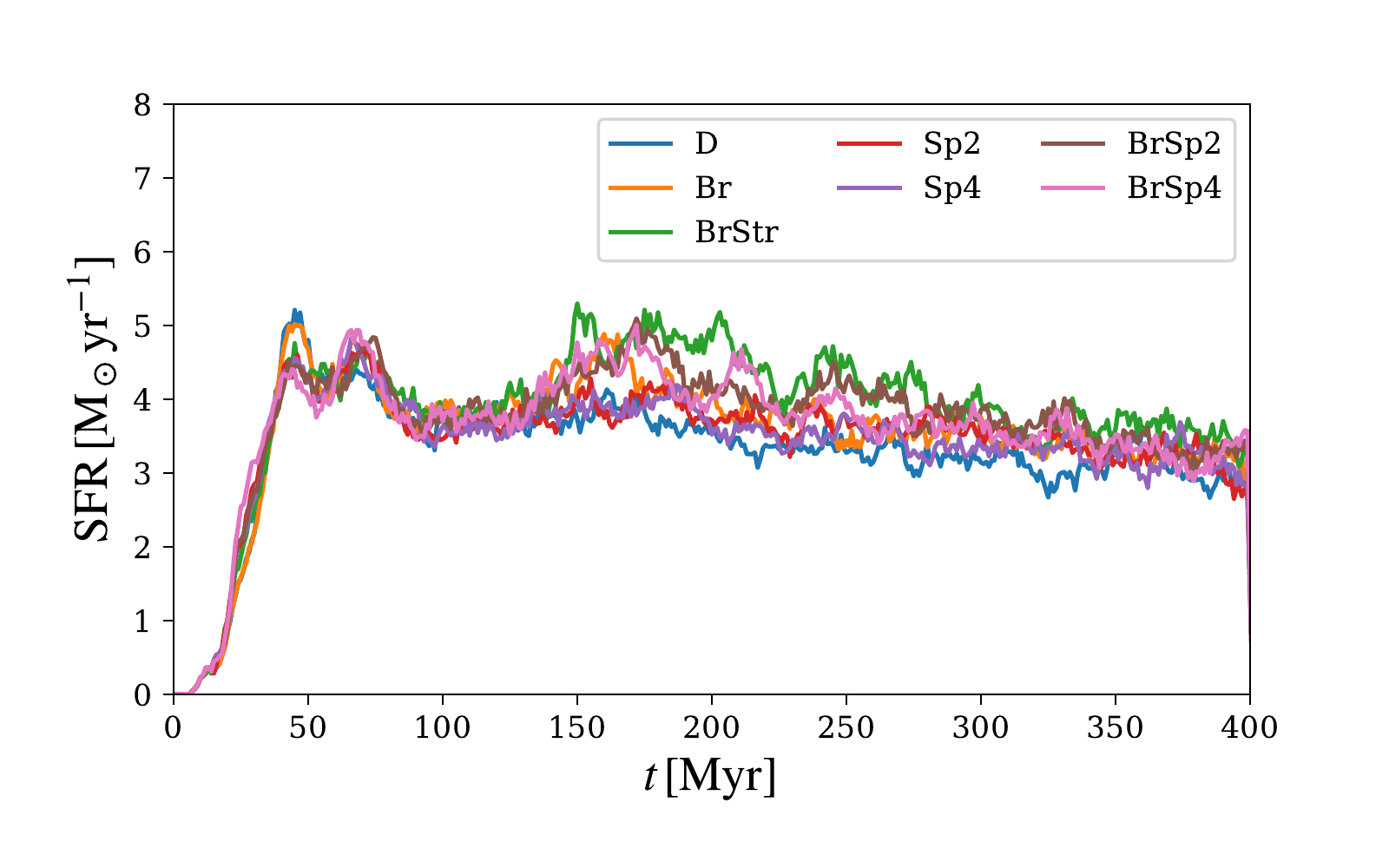}
 \caption{Total star formation history of each of the simulated discs. The global star forming history is effectively the same in all models.}
\label{sfr_global}
\end{figure}

It is worth noting that the BrSp2 model displays a 4-armed structure in gas and very young stars, but the old stellar population (represented here by spiral and bar potentials) only trace a 2-armed morphology. This is because the stellar response to rotating bars is not believed to drive any arm features, merely orbital families parallel and perpendicular to the bar, but gas clearly does \citep{2014PASA...31...35D}. This is could explain the apparent dichotomy between 2 and four arm models in the literature, where a strong 2-armed arms are seen emanating from the bar in some studies, whereas radio surveys tend to see a clearer 4-armed structure \citep{2001ApJ...556..181D,2008ASPC..387..375B,2016MNRAS.456.2885R}.

Maps of the young stellar material are shown in the left panels of Figure\;\ref{ImsAll_G}. The stars tend to trace out the densest regions of gas relatively well, though show less of the filamentary inter-arm features. The star formation histories for each disc are shown in Figure\;\ref{sfr_global}. While our total star formation rate is somewhat higher than observed for the Milky Way \citep{2016ARA&A..54..529B}, we are primarily interested in where stars are forming, not the absolute rate\footnote{This could easily be fine-tuned via the sub-grid physics models, though we deem this unnecessary for this work.}. Each model has effectively the same history, with an initial onset of star formation as the gas cools from the $1\times 10^4$\,K initial conditions, which progresses at a near constant level for the remainder of the simulation. The barred models show a slight increase once the bar reaches full strength, at around 170\,Myr. Even though models show a strong bar or bisymmetric spiral pattern, they will form stars at a similar rate.

We note that specific choices of subgrid physics could potentially impact these aforementioned features. Surprisingly little work has focused on how such recipes specifically impact specific features such as bars and spiral arms in the gas, with most simply comparing results with and without feedback included. Many studies do exist comparing feedback models in discs \citep{2017MNRAS.466...11R,2018MNRAS.478..302S}, though they mostly focus instead on the star forming potential and resulting scale-heights. Preliminary tests with the feedback model of \citet{2014MNRAS.442.3013K} did show some minor differences in gas response, with the main change being a slight decrease ($\sim 10$\%) in the total SFR systematically across the disc. The calculations from \citet{2014arXiv1406.4150P} contain the same potentials used in this study, but with different subgrid physics (the cooling rates differ, as does the addition of self-gravity, star formation, and feedback). The smaller scale features seem most affected by these additions, with the hydro+cooling only runs of \citet{2014arXiv1406.4150P} containing much more small scale bifurcations and branches in the smoother gas disc. For instance, the Sp2 and simulations in this paper show a clear 2-armed structure, and only a slight indication of branching features between arms, which are much clearer in the runs of \citet{2014arXiv1406.4150P}. Being secondary arm features, these seem more sensitive to disruption by feedback, which dominates the morphology of the smaller scale ISM. The mixed bar-spiral models also display very similar morphologies, with the differing pattern speeds giving rise to time-dependent structures. Our Sp2 and Sp4 models appear very similar to those of \citet{2011MNRAS.417.1318D,2013MNRAS.432..653D}, who utilise the same spiral potential (4 and 2 armed respectively) but with a different numerical code and subgrid physics routines. There is a clear need for further work on the impact of subgrid physics on the detailed gas response to spiral and bar structures.

\subsection{Structures traced by different stellar populations}

\subsubsection{The distribution of ages at a given instance}

\begin{figure*}
\includegraphics[trim = 10mm 20mm 20mm 10mm,width=180mm]{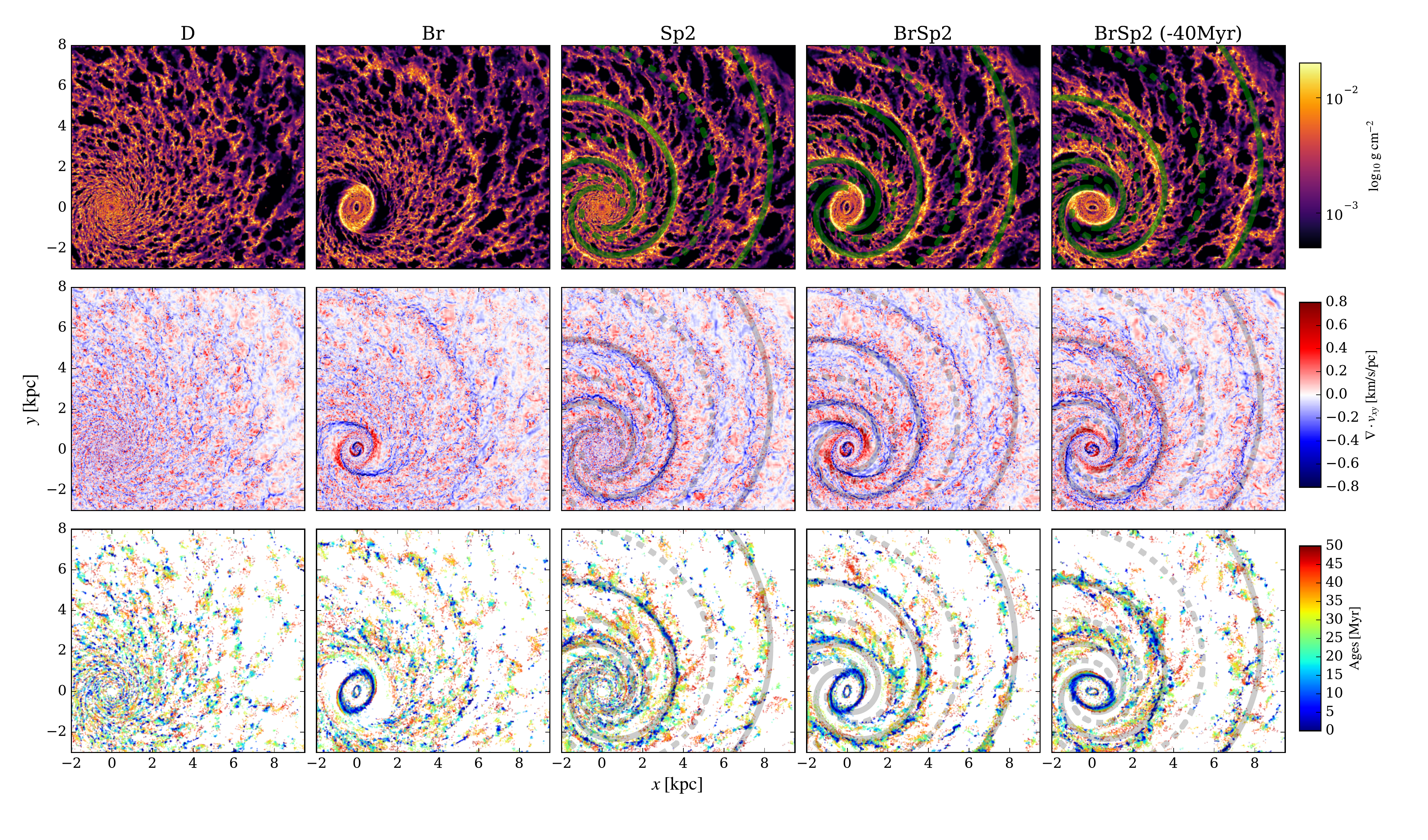} 
\caption{
Top row: gas density in a number of galactic simulations. Middle row: the divergence of the in-plane gas velocity field, with blue regions indicating converging gas flows. Bottom row: the locations of young star particles, coloured by age. Arm minima are shown by solid lines, with dashed lines indicating the mid-point between minima. The right-most column is an additional snapshot of BrSp2, orientated so the spiral potential is aligned in agreement with the other panels.}
\label{sf_events}
\end{figure*}

The star formation prescription allows us to directly trace where star formation events occur in our simulations, and due to the potentials representing the arms and bar, we also know precisely the location of the potential minima at any given time. In Figure\;\ref{sf_events} we show a zoom in of gas and young star particles in several simulations. The top row shows the gas density in a simulations of the axisymmetric disc, barred and two-armed models. An additional snapshot of the barred-spiral model is shown 40Myr prior to the other snapshot, re-orientated so that the spiral potential is at the same position as the other panels (spiral minima shown as solid lines, inter-arm position as dashed lines). The second row shows the divergence of the in-plane velocity of the gas particles (\delv{}), with a converging gas flow being a criteria for the star formation prescription. The bottom row shows the locations of the younger star particles in the simulation ($<50$\,Myr), colour-coded by their ages.

Star formation in the D model appears uniformly distributed, though clumpy/filamentary on the smaller scales. The velocity divergence and gas density are similarly uniform across the disc. The barred and spiral models have a clear preference for star formation to be re-distributed in non-axisymmetric features. The Br model gathers gas into arm-like features in the mid-disc, creating ridges of a converging gas flow that allows for concentrated star formation. In Sp2 the youngest stellar population is strongly associated with the spiral arm potential. Most of the blue-coloured particles (those that have just formed) trace the spiral arms, where the dense gas is located, which is also well associated with regions of \delv{}$<0$. There does remain pockets of star formation off-arm, for instance the branch lying between the two arms around $x=5$\,kpc, lying very near to the dashed lines. Such branches/bifurcations are believed to be caused by secondary shocking of the gas near the ultraharmonic (4:1) resonance for sufficiently strong spiral forcing \citep{1973ApJ...183..819S,1994A&A...286...46P}, also showing a very slight over-abundance of converging gas flows. In Sec. \ref{sec:model} we present a more detailed discussion of the gas morphological response to such perturbing potentials with a simple analytic model.

The highest concentrations of young stars do not trace the potential perfectly. The inner-disc of Sp2 clearly shows the star formation front downstream of the spiral potential (see $x=0$kpc, y=$-2$kpc), and slightly upstream in the outer disc (note co-rotation lies just beyond the disc edge in these simulations). This complex relation between the shocked dense gas and spiral minima has also been observed in previous theoretical studies \citep{1969ApJ...158..123R,1973ApJ...183..819S} and simulations \citep{2004MNRAS.349..909G,2014MNRAS.440..208K,2015PASJ...67L...4B}, with the radial dependence of the mach number playing a key role in where the downstream--upstream offset transition occurs. 

\begin{figure*}
\includegraphics[trim = 0mm 0mm 0mm 0mm,width=180mm]{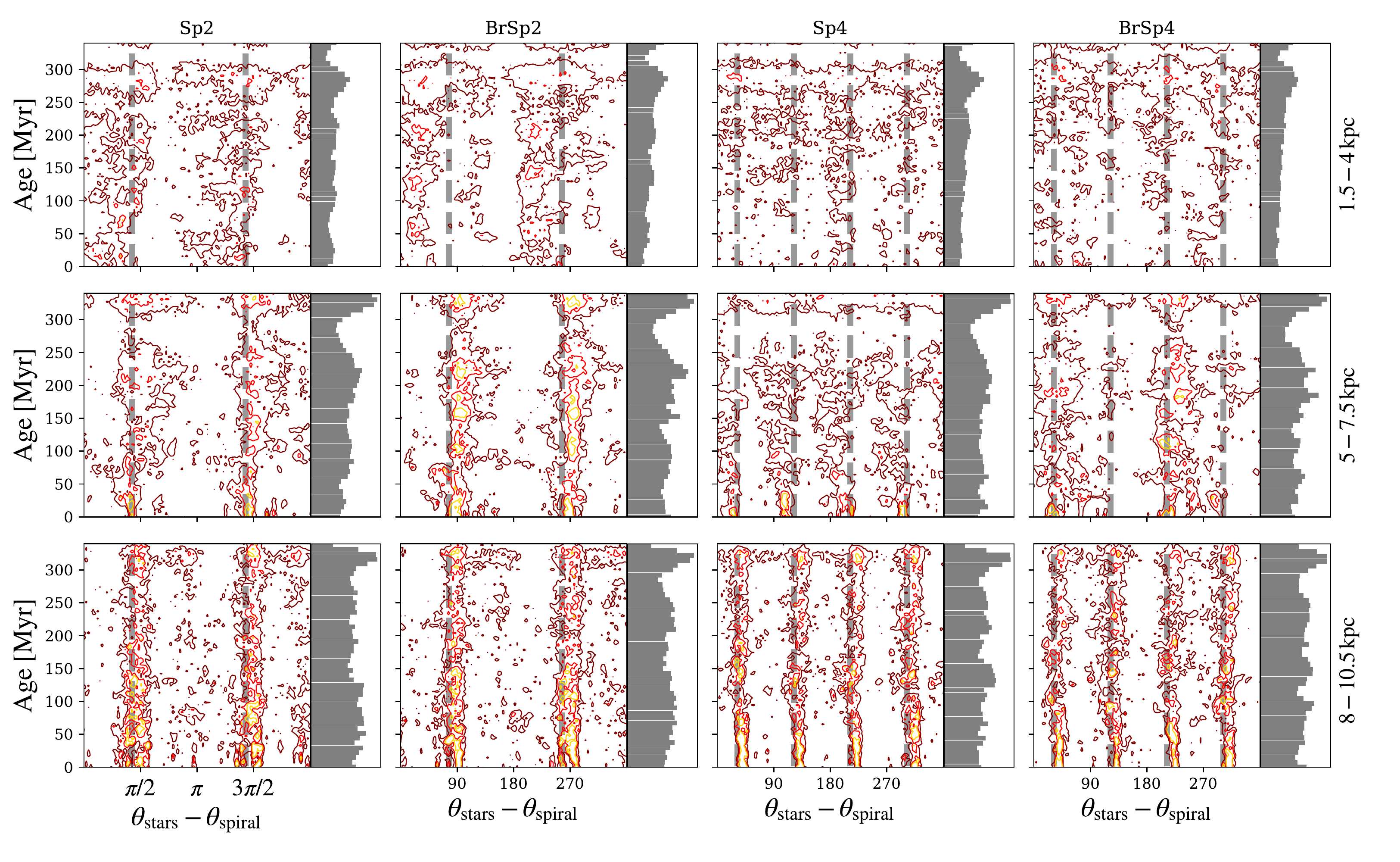} 
\caption{The azimuthal position of stars of different ages in all armed simulations at 400\,Myr. Azimuthal positions are plotted with respect to the imposed spiral arm potentials. Each row shows a different radial range: inner, mid, and outer disc in the top, middle and bottom rows respectively. On the right hand side is the same data collapsed down the azimuth axis as a 1D histogram.}
\label{ageVtheta}
\end{figure*}

While the locations of stars with respect to the spiral arms is still relatively unsurprising in Figure\;\ref{sf_events}, the addition of a bar in the centre region clearly acts to change the global structure of dense gas. The addition of the bar to the spiral model acts to alter the locations of star forming events. The two epochs of BrSp2 shown here highlight when the bar-driven spiral arms are midway between the Sp2 arms (fourth column) and when the are coincident (fifth column). At the fiducial time-frame the dense gas is shifted considerably compared to the spiral minima around $x=4$\,kpc (also evident by the locations of low \delv{} and the young stellar population). 40Myr prior to this (fifth column) the bar is orientated $+90^\circ$ compared to the other maps, at which point the peak density and young stellar population lies upstream of the arm located at $x=4$kpc. At this time both the converging flow and gas density peak are shifted upstream of the plots in the fourth column, directly impacting the location of the youngest stars in the lower panels. Interesting, in both panels the converging velocity flows trace and unbroken line into the galactic centre and inner gas disc, with no clear break as the disc evolves, despite the bar rotating much faster.

A clearer and more complete picture of where the young stars are with respect to the spiral arms is shown in Figure\;\ref{ageVtheta}. Contours indicate the density of star particles of a particular age at a given azimuth, which has been phase-shifted so that the bottom of the spiral potential traces a vertical line (the grey dashed line). Each column represents a different model that includes the spiral potential, and the plots are created at the end of the simulations (400\,Myr). Each row is a different radial annulus of the disc. The oldest stellar material is at the top of each panel, with the youngest stars at the bottom. 

Sp2 tells a similar story as shown in Figure\;\ref{sf_events}, with the younger stars tracing a much narrower range of azimuth. The inner-disc shows the clear offset in the young stellar material to be downstream of the potential well (smaller values of $\theta$). Several horizontal features here and in the mid-disc correspond to the small spur-like features in the younger population (Fig.\;\ref{ageVtheta} top-left panel, the horizontal patch of $<50$Myr stars emanating from the arm around $[\pi,3\pi/2]$). In the outer disc (bottom left panel) the interarm regions also correspond with the branching features, seen as a low density feature almost equidistant between the spiral arms. With the addition of a bar, there is not a strong difference in the overall structure of these stellar count maps in Fig.\;\ref{ageVtheta}, but their location is shifted in azimuth due to the rotation of the bar acting to pull the stars away from the spiral minima in the inner-disc, drawing it further downstream of the arm. This effect is time dependant, with the peaks in the stellar material moving from upstream to downstream, depending on the bar orientation (this will be discussed later in this section). The bar also acts to strengthen the arms in the mid-disc compared to the un-barred model (contours are of the same levels throughout the figure). 

For the BrSp2 model there are over-densities along each arm at nearly regular intervals (in the inner and mid-disc where both arm and bar features are present). For instance, in the inner-disc panel, there are blobs at ages of approximately 50, 120, 200 and 280\,Myr. The relative rotation frequency between the arms and bars is 40\,\ps{}, which corresponds to a relative time offset of 153\,Myr, and as the both components are bisymmetric, that means the bar will pass the an arm every $\sim77$\,Myr, which is approximately the spacing between these ``beats" for BrSp2. That is; small boosts in the young stellar population are seen to correspond to the over-lap frequency of the arms and bar, corresponding to when the combined potential minimum is the deepest. The effect is still rather subtle, but the vertical histograms all particles binned by age do show alternating peaks in the mid-disc in particular for BrSp2 and BrSp4.

For the 4-armed models, the gas response is less clear. For Sp4, the younger stars trace the spiral potential fairly well, especially in the outer disc where the gas response is strongest. The youngest stars also show different offsets compared to the potential well, with the mid-disc showing an downstream offset and the outer disc showing a upstream one. The BrSp4 model shows different features. Stellar material is also bunched closely together, with vacant regions at $180^\circ$ and $360^\circ$, whereas the other interarm regions host more material. This can also be clearly seen in the top-down map of Fig.\;\ref{ImsAll_G}, where two minor arms rest on the convex side of the major arms that are reinforced by the bar, with a wide dearth of stars downstream. There is seemingly a change in arm number, with the mid-disc showing a two-arm morphology, branching out into four-arms in the outer disc. As with BrSp2, this appears to be a result of the bar aligning with two or the arms in particular. The beats alluded to in the BrSp2 model are much harder to see here in the top contours of Fig.\,\ref{ageVtheta}, but can be seen in the vertical histogram in the mid-disc panels, approximately spaced out as they would be expected (now only $\sim 38$\,Myr apart due to the 4-fold symmetry).

\subsubsection{The impact of differing pattern speeds}

\begin{figure}
\includegraphics[trim = 0mm 0mm 0mm 0mm,width=80mm]{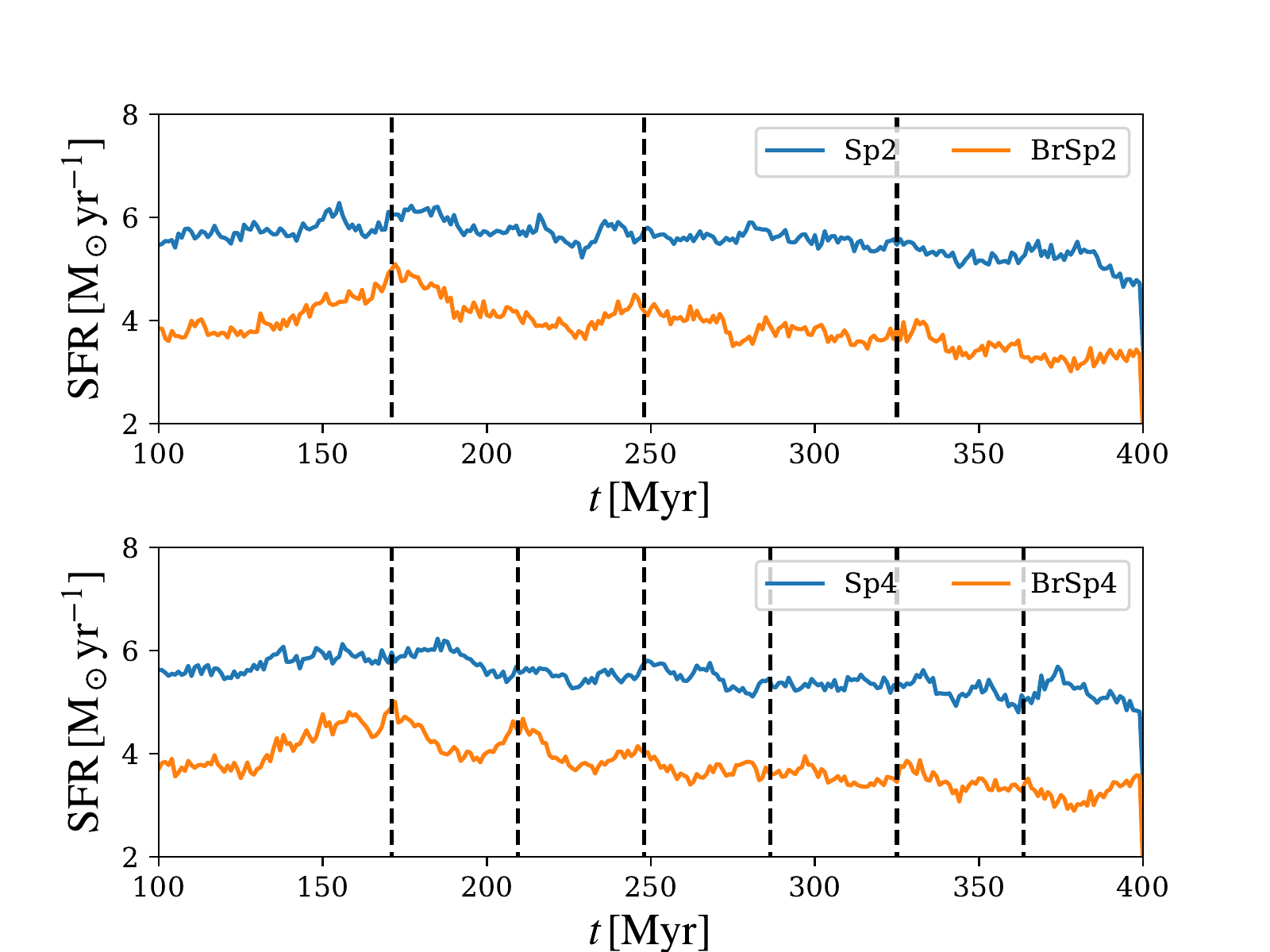}
 \caption{Star formation rate as a function of time for all spiral models, with ``beat" frequencies shown as vertical dashed lines. The unbarred SFR are offset vertically by $2M_\odot {\rm yr^{-1}}$ for clarity.}
\label{SFRbeat}
\end{figure}

\begin{figure*}
\includegraphics[trim = 0mm 0mm 0mm 0mm,width=150mm]{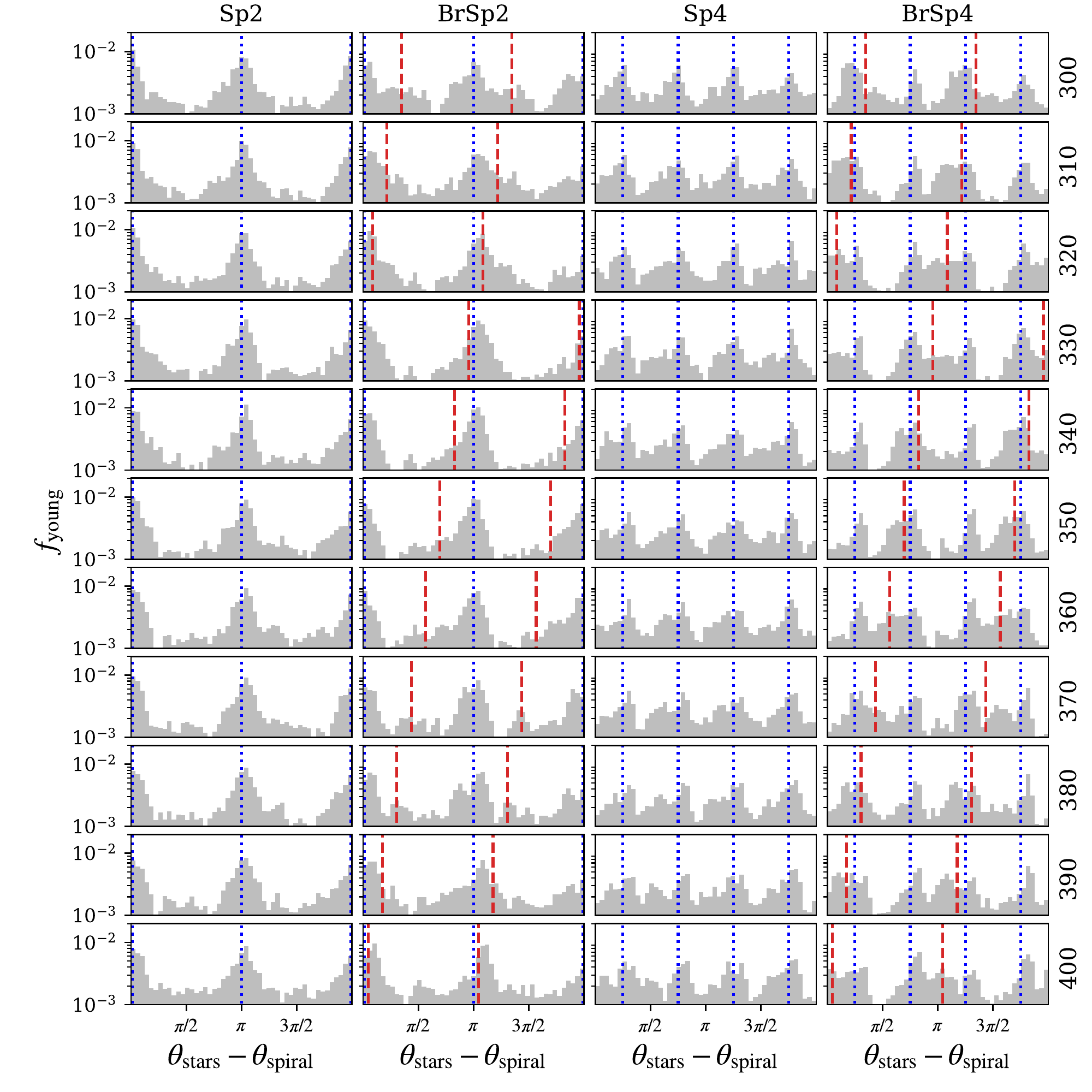}
 \caption{Young stars (less than 50\,Myr old) at radii $>2.5$\,kpc placed in azimuthal bins phase-shifted such that the $x$-axis is the offset with respect to the arm potential. Each row indicates a different time in the simulation (300--400\,Myr). The dashed blue vertical lines indicate the arm potential, and the red lines the bar potential which moves to the left in the rest-frame of the arms due to faster rotation speed.}
\label{SFbeats}
\end{figure*}

The existence of a beat-frequency that plays a role in star formation is clear when the SFRs in Fig.\;\ref{sfr_global} are looked at more closely. Figure\;\ref{SFRbeat} shows the total star formation rate, summer over all radii and azimuth, for all armed models. Sp2 is shown in the top panel and Sp4 in the lower, with the unbarred models artificially offset vertically by $2M_\odot {\rm yr^{-1}}$ for clarity. Vertical black dashed lines show the beat frequency, i.e. the times between the spiral and bar-arms overlapping, their horizontal zero-point offset simply chosen to place the first vertical line at a time where an overlap between feature is seen by-eye (as while we know the bar/arm positions at all times, the gas response to the bar that generates the supplementary arms is non-trivial). There is a clear correspondence between the spacings of the vertical lines and peaks in SFR in the BrSp models that is not present in the unbarred ones. This becomes less pronounced with time as the gas disc is depleted, though there are still clear troughs between peaks at later times (400\,Myr is just prior to the next overlap, and both BrSp2 and BrSp4 are rising).

As the bars appear to drive a time-dependence in the measured arm locations in young stars, we plot similar data but across multiple time-frames in Figure\;\ref{SFbeats}. We again show the azimuthal locations of stars (plotting the fraction of young stars, $f_{\rm young}$, at a given azimuth), though this time only showing a younger stellar subset, phase shifted to the arm potential. Each column shows a different armed model, and each row a different time in the simulation. We define young stars as those with ages $<50$\,Myr, thus each row in Fig.\,\ref{SFbeats} shows a different selection of stars (e.g. the 350\,Myr row shows stars born between 350--300\,Myr). Blue dotted lines show the minima of the arm potential. Red dashed lines show the bar orientation with respect to the spiral arms. This acts as an indicator of the impact of the bar rotation, but note that arms driven by the bar will clearly not lie along the red dashed line, instead spanning a range of azimuth as radius increases, and also being more tightly wrapped than logarithmic spirals.

The Sp2 and Sp4 models in Figure\;\ref{SFbeats} show a relatively simple picture, with young stars distributed evenly around the minima of the spiral potential. While offsets clearly exist, as shown in the previous section, they average out over the disc so that star forming regions lie uniformly around the potential minimum. The barred models show additional peaks in the distribution that move in phase with the bar orientation. For example, in BrSp2, there are minor peaks that trace the red dashed lines through azimuth space, and also causing a skew in the distribution of the primary peaks when they are just off-centre of the spiral minima. This effect is even clearer in the BrSp4 simulations, where the bar phase is clearly associated with an additional stellar peak, causing either a clear bridging of stellar material between arms (e.g., 320\,Myr and 400\,Myr) or reinforcing the strength of two of the four arms (350\,Myr, 310\,Myr).

This, as well as the data in Figure\;\ref{ageVtheta}, show that an out of phase bar can clearly create additional arm-like features in the young stellar population/gas that may not be present in the old stellar population. It can also cause two out of four arms to appear stronger when the spiral and bar-driven arms overlap, as well as create inter-arm features between a two-armed spiral density wave. This is intrinsically tied to a beat--like feature in the star formation rate, caused by the superposition of arm and bar features.

In Appendix\;\ref{Appx2} we show similar analysis for combinations of the spiral potential and the stronger, slower rotating, bar. The trends and features are very similar to those described above, and are to be expected for any combination of rigidly rotating bars and spirals with differing pattern speeds such that the region between their and ILR and OLR overlap.

\section{Observational analogues}
\label{sec:discussion}
\subsection{Locations of star formation}
\begin{figure*}
\includegraphics[trim = 0mm 0mm 0mm 0mm,width=180mm]{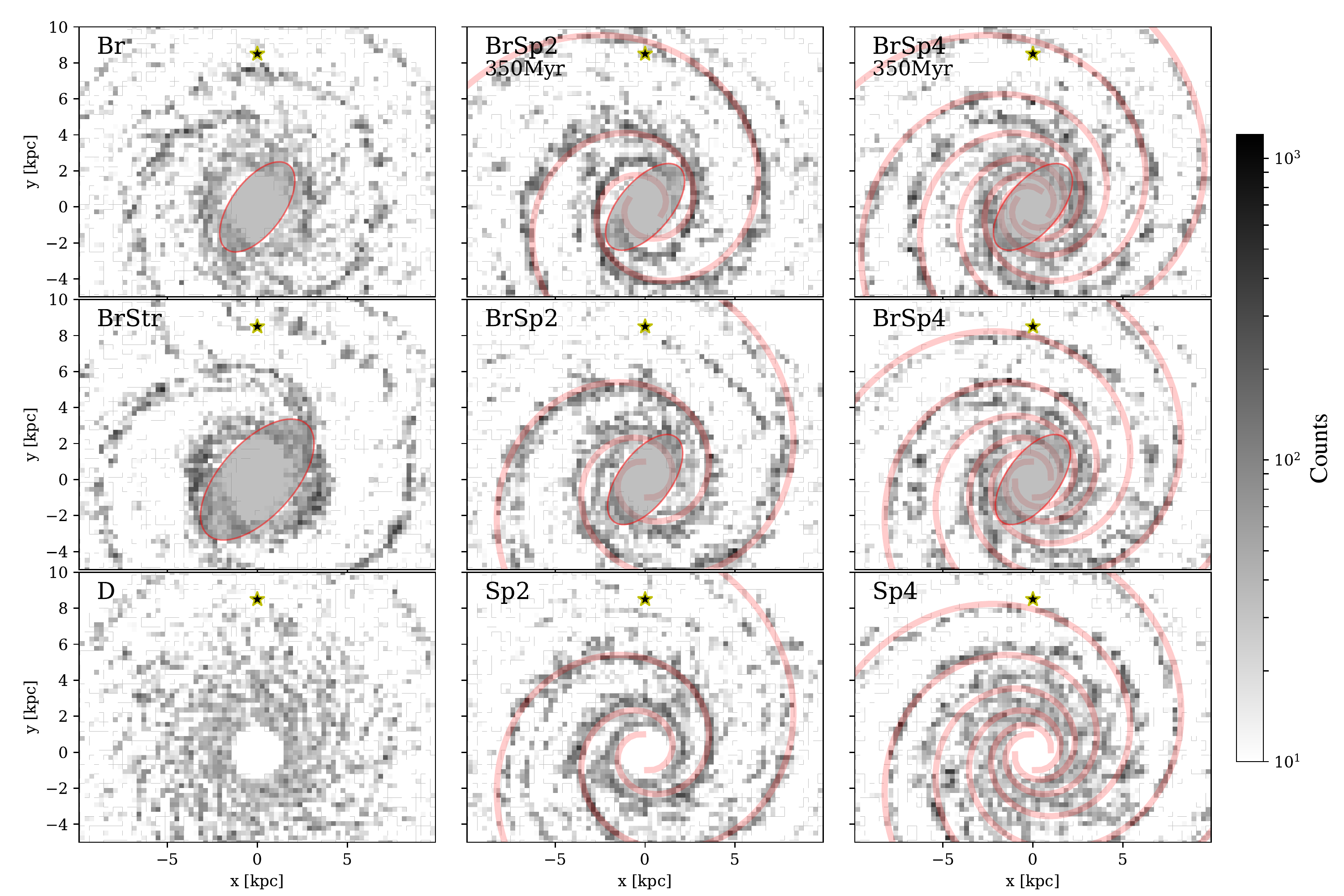}
\caption{Binned star formation events in the simulated galaxies. Simulation data is simply binned locations of young ($<50$\,Myr) star particles. If all star particles have approximately the mean stellar particle mass (1400$\rm M_\odot$) then a conversion factor for our cell size of (0.26kpc)$^2$ is $4\times 10^{-4}{\rm M_\odot \, kpc^{-2}yr^{-1}}$ that will convert counts into a $\Sigma_{\rm SFR}$. Each galaxy is shown at 400\,Myr, and the mixed spiral-bar models are shown at an additional time-stamp of 350\,Myr. Red lines indicate the minimum of the spiral potential, and the grey ellipse the orientation of the inner bar. The star symbol corresponds to the Solar position.}
\label{MySFGAL}
\end{figure*}

\begin{figure}
\includegraphics[trim = 0mm 0mm 0mm 10mm,width=80mm]{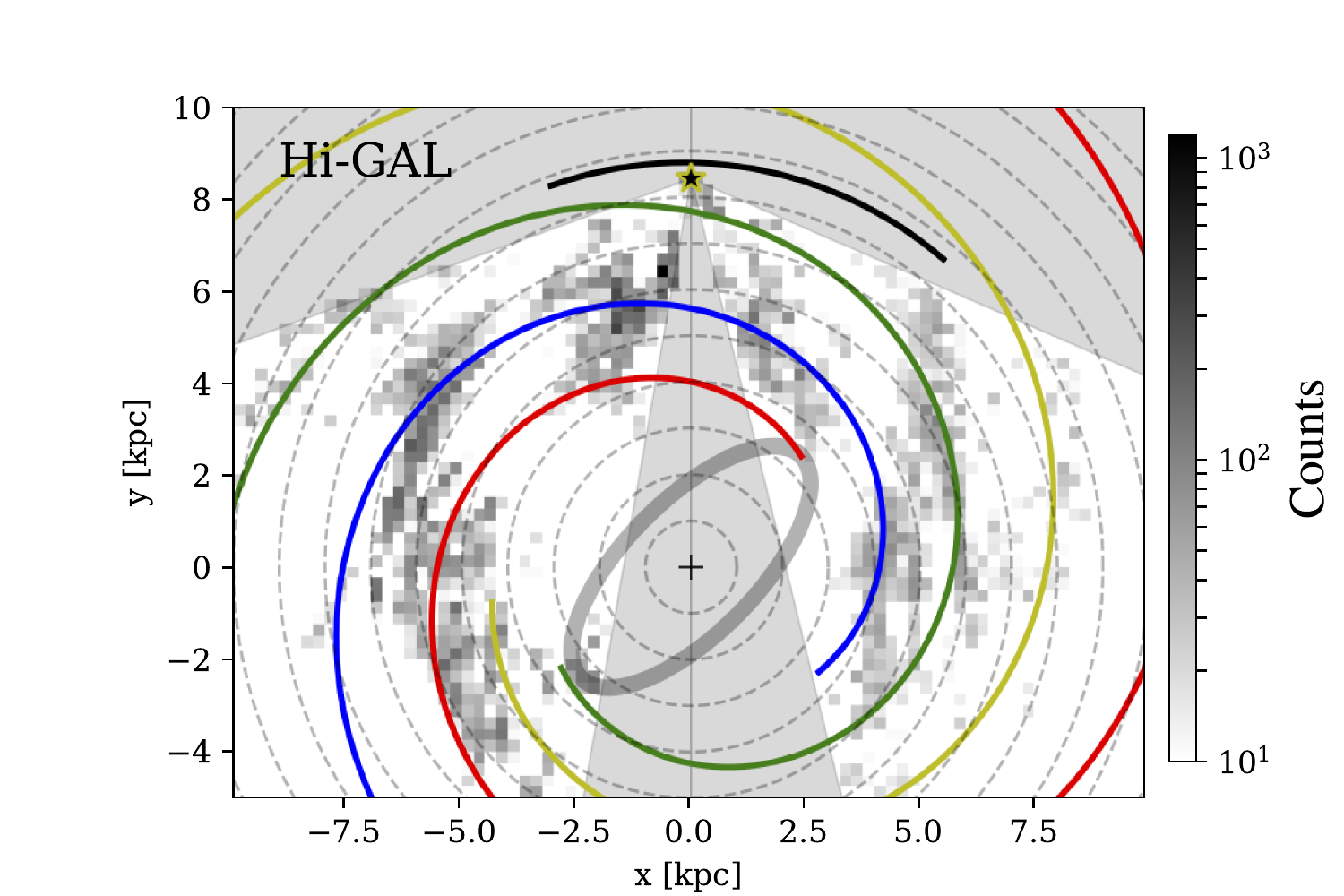}
 \caption{Star forming regions from the Hi-GAL survey, with coloured spiral arms from \citealt{2014A&A...569A.125H}. Shaded regions show exclusion region of the survey.}
\label{HGSFGAL}
\end{figure}

We will now compare the results of these simulations to observational diagnostics of Galactic structure and star formation. Figure\;\ref{HGSFGAL} shows a 2D histogram of the locations of  compact clumps from the Hi-GAL survey \citep{2010A&A...518L.100M,2016A&A...591A.149M,2017MNRAS.471..100E}, including both star forming and quiescent sources. Spiral arms of \citet{2014A&A...569A.125H} are over-plotted as coloured lines, based on locations of  H\,{\sevensize II} regions, where red is the Norma arm, yellow the Perseus arm, green the Sagittarius-Carina arm, blue the Scutum-Centaurus arm and black the local arm. While the disc is only partially covered (survey extremities and kinematic distance uncertainties in the Galactic centre limit the data to $15^\circ<|l|<70^\circ$, excluding shaded regions in the figure), there is a correspondence between the spiral arms and the locations with highest counts, but it is by no means a 1:1 tracing. Some caveats to keep in mind are that uncertainties in positions are of the order of 30\%, and the number of nearby clumps is more complete due to the mass sensitivity and the resolution being poorer for more distant sources.

The panels of the Figure\;\ref{MySFGAL} show the locations of star formation within the simulations. For all the results of this section the Sun is placed at a distance of 8.5kpc, and the bar is angled at 45\arcdeg{} to the Sun-Galactic centre line. No attempt is made to re-orientate the arms to similar locations to those seen in the Milky Way, which would require selecting a specific time period in the simulation with a specific bar-arm offset, and so results are not expected to perfectly reproduce Galactic features.
Star forming regions are defined as those harbouring star particles younger than 50\,Myr. Over-plotted are the positions of the spiral arm and bar potentials. The locations of high counts are clearly correlated to the arm potentials when present. For the D model, the star formation is effectively uniform across the disc, and in  Br/BrStr there is a clear alignment along the bar-driven arms. The Sp4 model has a fairly clear-cut structure, with inter-arm star formation becoming more prevalent within the ILR at $\sim 7$\,kpc (though still mainly concentrated along arms until the inner 2:1 resonance at around 3kpc). The Sp2 model has a considerable amount of inter-arm structure, with well defined branches that lie between the primary spiral arms. There are also hints of star forming regions tracing out spur/feather features in the inner-disc. For instance, at around $(4,-2)$kpc and there are clear spurs pointing near-perpendicular to the arms. Such star forming spurs are seen in other disc galaxies, most noticeably in M51, where \citet{2017ApJ...836...62S} show that multiple tracers of star formation are concentrated in patches in the spurs of the arms.
 
\begin{figure*}
\includegraphics[trim = 0mm 0mm 0mm 0mm,height=200mm]{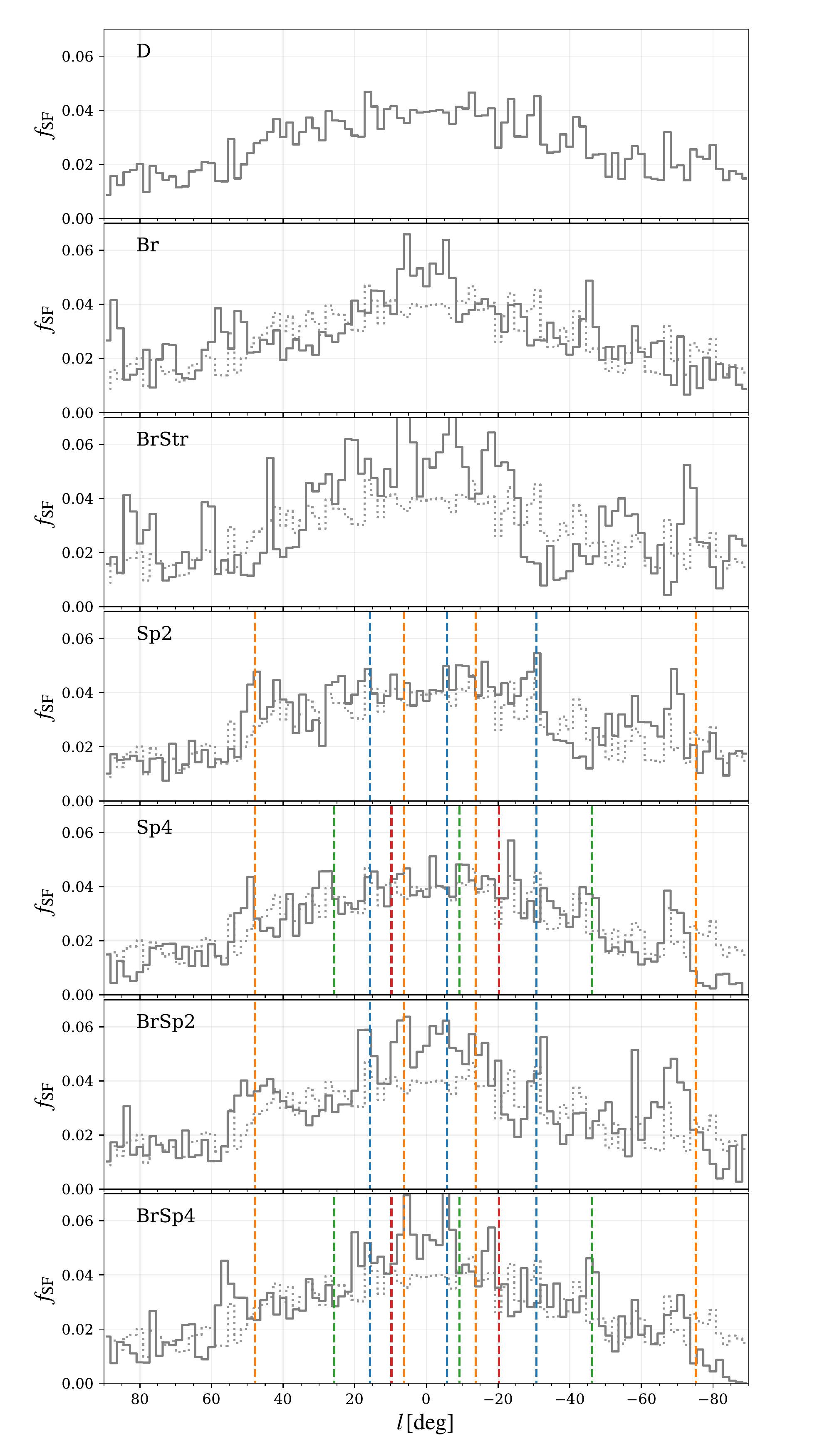}
\includegraphics[trim = 10mm 0mm -10mm 0mm,height=200mm]{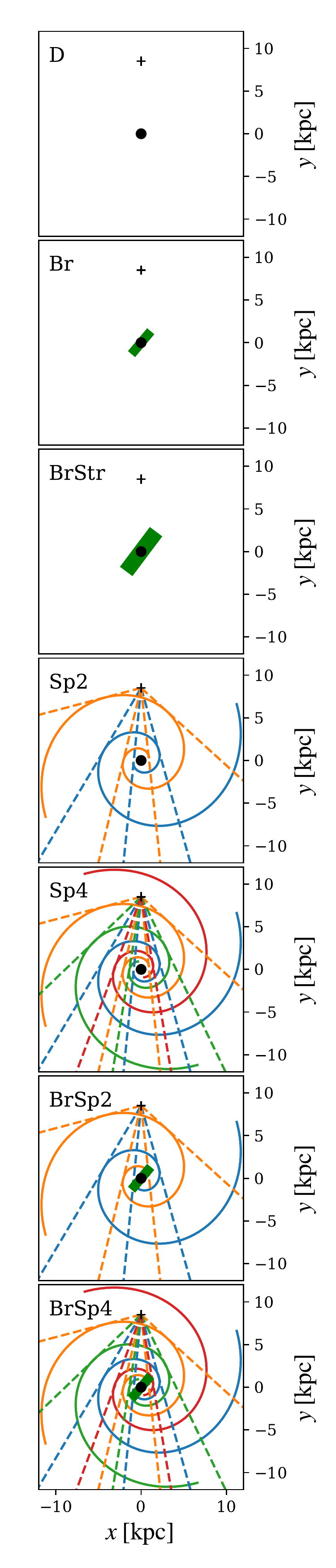}
\caption{Left: ratio of the amount of young stars (ages $<50$\,Myr) verses gas as function of longitude in 7 different Milky Way simulations. Arm tangents are plotted in each panel where applicable. Right: top down maps displaying the position of the non-axisymmetric features in the old stellar population that drives gas over-densities, with arm tangents plotted using the same colour scheme as the plots on the left. Dotted histograms show the data from the D model in each other panel. The Sun is placed at a distance of 8.5kpc, and the bar is angled at $45^\circ$.}
\label{SFlong}
\end{figure*}

\begin{figure}
\includegraphics[trim = 15mm 0mm 5mm 0mm,width=90mm]{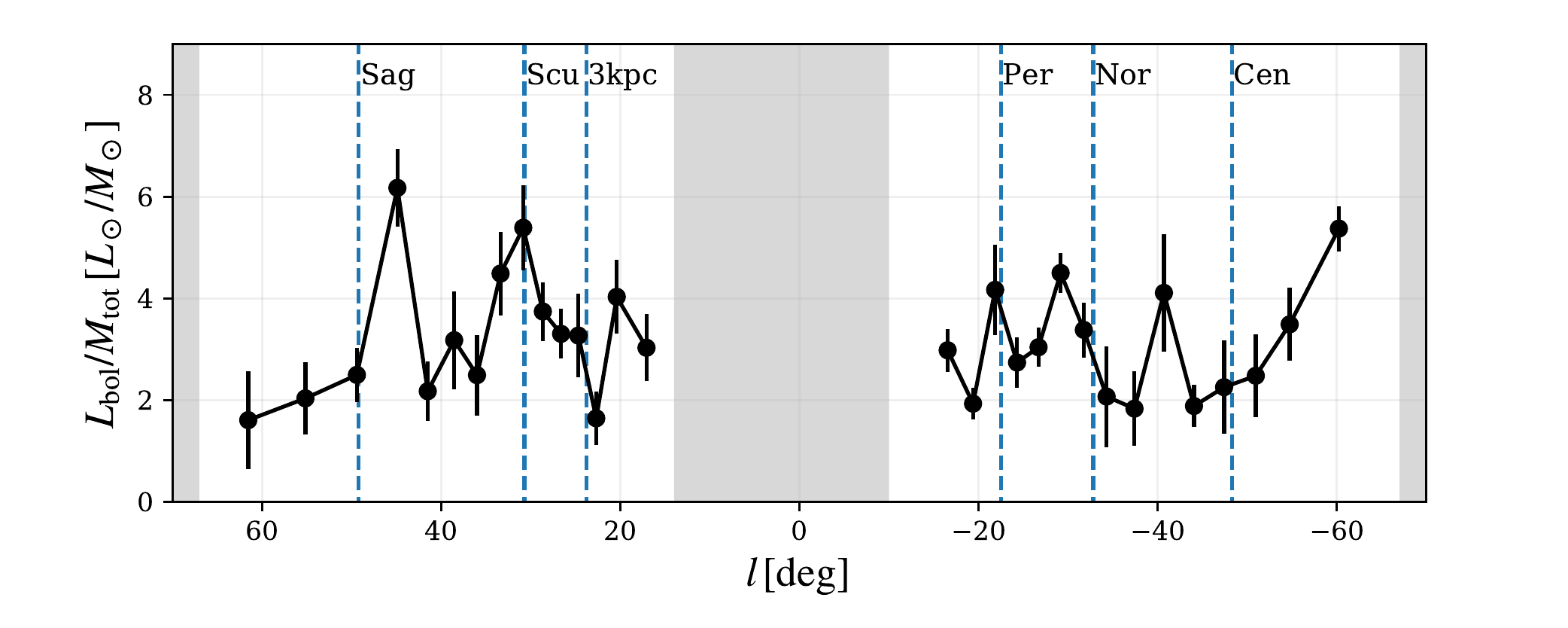}
 \caption{Azimuthally binned luminosity ($L_\mathrm{bol}$) to mass ratio for Hi-GAL sources along difference lines-of-sight in the inner Galaxy. Data is binned into equally populated bins, rather than equal sized ones. See \citet{2018MNRAS.479.2361R} for further details. Shaded regions denote those outside the survey area. The tangencies for Sagittarius, Scutum, 3kpc (near), Perseus, Norma and Centaurus arms from \citet{2015MNRAS.454..626H} are plotted as dashed vertical lines.}
\label{HigalLong}
\end{figure}

As the relative orientation of the arms and bars is time-dependent in the BrSp models, we show the star forming regions at two different time frames; 400\,Myr and 350\,Myr. Looking at the BrSp2 models there are clear differences between the two time-frames. While there are inter-arm star forming regions in both cases, their location with respect to the spiral arms is different. At 400\,Myr the inter-arm material is concentrated on the concave side of the spiral (upstream of the arm) whereas at 350\,Myr it is primarily on the convex side (downstream of the arm). The 350\,Myr timeframe displays no clear interarm branch (as seen in BrSp2 and Sp2 at 400\,Myr), as the bar has effectively swept it into the primary spiral arms. The BrSp4 model shows less differences to the Sp4 model, which is to be expected as the main resonant features of the four-armed potential are at larger radii compared to Sp2. Note that Fig.\,\ref{SFbeats} implies that while 350 and 300\,Myr snapshots do have a different arm-bar phase offset, this should be a maximum at 370\,Myr. While we do not show the map here, it clearly shows spiral arm features not associated with the $m=2$ potential, as seen in the bottom left of Fig.\,\ref{ImsAll_G}.

There is a particular feature of note in the BrSp4 model at 400\,Myr. Looking at the star formation in the mid-disc around $x=\pm5$\,kpc and $y=\mp 2$kpc, two of the arms have very little associated star formation. This is mirrored in the Hi-GAL data, where is a dearth of star formation in the Norma arm (red) at $(-5,2)$kpc, though the feature is not as clear in the Sagittarius-Carina arm (green), though this does lie behind the bar end in the partial shadow of the Galactic centre. Star forming regions near this location in the simulation are instead located in the inter-arm region rather than tracing the arms themselves, joining as a branch to the next arm. This is similarly seen in the Hi-GAL data between the Norma (red) and Scutum-Crux arm (blue). At this radii in the simulation the gas response to the four-armed potential is fairly weak (see the Sp4 map), and so the bar response is what is driving the flow of gas, which is causing these arm sections to ``peel away" from the underlying spiral potential.
 
To re-cast this data into a more observable-like manner we plot the line-of-sight distribution of star formation regions in our simulations in Figure\;\ref{SFlong}. This figure shows locations of young star particles (those with ages $<50$Myr) binned by Galactic longitude as viewed from the Sun's location, taken to be $y=8.5$\,kpc. Data is binned by $f_{\rm SF}$: the ratio of mass in young stars divided by gas mass along a given line of sight. The data is shown for each model at 370\,Myr, chosen for when the bar is orientated 45$^\circ$ to the Sun--Galactic centre line and spiral arms lie in regions approximately corresponding to their measured locations. This bar angle is suggested by some works \citep{2011ApJ...733...27G}, though a somewhat larger angle than suggested by others \citep{2016ARA&A..54..529B}. However, out goal here is to show the impact of the bar--spiral combination, rather than match the observed morphology exactly. In the right column a top-down schematic view of the galaxy is shown, with the Sun shown as the black cross, and the Galactic centre as the black circle. Spiral arms and bars are indicated by coloured lines, with their orientation corresponding with the vertical lines in the left panels. Dotted lines indicate the spiral arm tangents, which are often used as observational tracers of spiral arms \citep{2000A&A...358L..13D,2010ApJ...722.1460S}, and these tangencies are over-plotted on the count maps in the left column.

These maps show a general agreement for peaks corresponding to the spiral arm tangencies, though there is significant noise in the interarm regions. For instance, in Sp2 the arm tangencies for $|l|<20^\circ$ seem on the same level as what is seen in the axisymmetric mode (D). There are a few regions where the peaks are offset from the spiral arms, such as the arm near $-75^\circ$ in each spiral arm model. This generally agrees with what is seen in the previous figures; that star forming regions are not necessarily exactly coincident with arm potential minima.

The impact of the bar can clearly be seen in Br and BrStr models, with clear peaks corresponding to bar-driven arms in $f_{\rm SF}$. These features are much clearer in BrStr than Br. The stronger bar also shows an over-density at $l=30^\circ$ which is not seen at $l=-30^\circ$, resulting from looking along the tangency of an arm beginning at the bar end in the first quadrant (see Fig.\;\ref{ImsAll_G}). While such bar-driven arms may be visible in observations, they would be identical to anything driven by spiral arm features. The inner tangencies of BrSp2 appear stronger than their its unbarred counterpart ($||l<30$\arcdeg{}). At this epoch the bar driven arms are coincident with the inner spiral arms of Sp2, but wind up faster at outer radii and so have less of an impact on the outer tangencies. See Sec.\;\ref{sec:model} for a more in depth look at the different arm types.

In Figure\;\ref{HigalLong} we show the luminosity to mass ratio of compact clumps in the Hi-GAL survey, as presented in \citet{2018MNRAS.479.2361R}, which serves as an observational analogue to the data in Fig\;\ref{SFlong}. Peaks are loosely correlated to the location of supposed spiral arm tangents (e.g. Norma near $-30^\circ$, Scutum near $+30^\circ$), though some arms such as Sagittarius ($l\approx 50^\circ$) are either strongly displaced to the tangent locations measured in other tracers, or non-existent. In the simulated data we also see several instances of peaks in $f_{\rm SF}$ displaced from the arm minima (a proxy for old stellar tracers), especially at $|l|\geq 45^\circ$ (e.g. BrSp4 near $l\approx 55^\circ$).

\begin{figure}
\includegraphics[trim = 0mm 0mm 0mm 0mm,width=80mm]{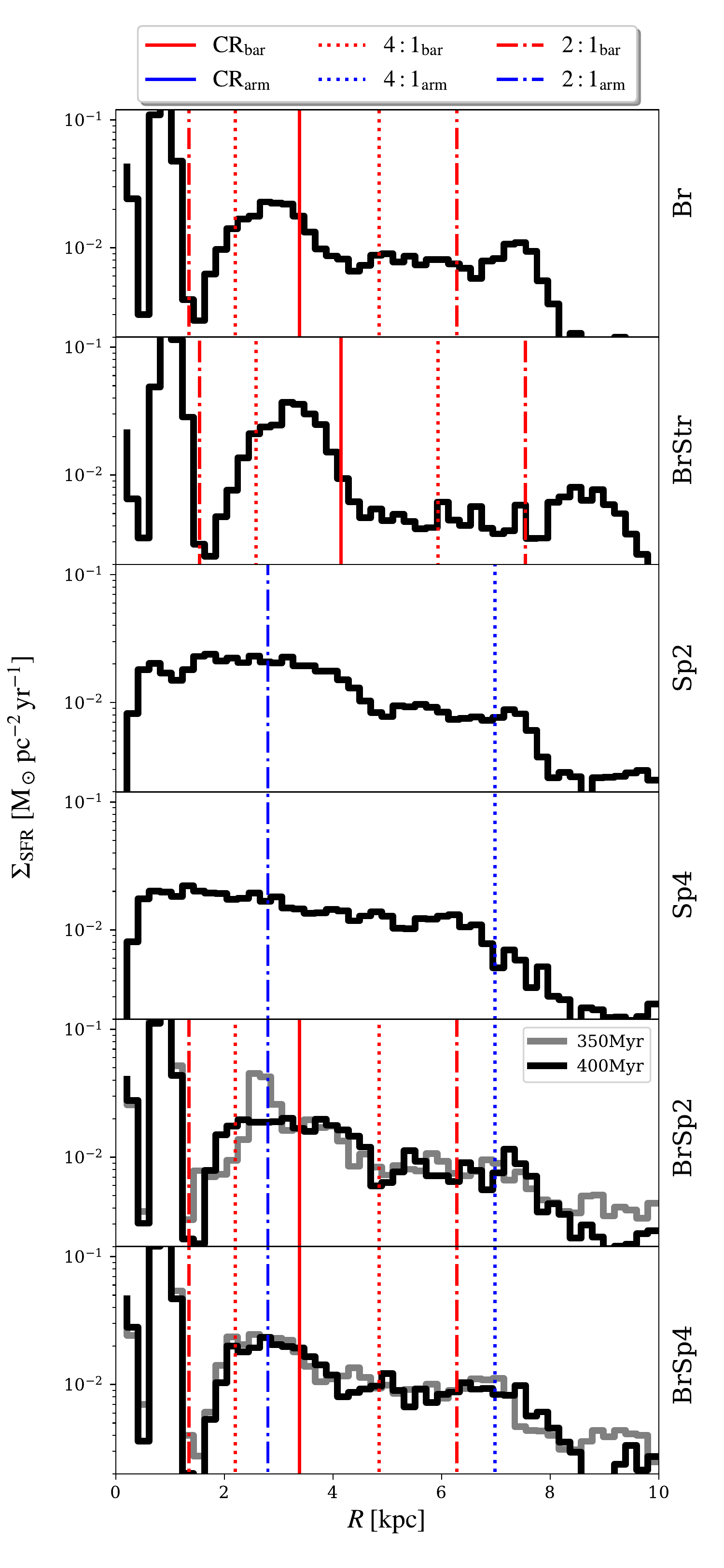}
 \caption{Radial distribution of star formation (traced by star particles with ages $<50$\,Myr) as a function of radius for models with imposed bar/arm potentials. $\Sigma_{\rm SFR}$ is evaluated in annuli of fixed width. The resonant radii shown in Fig.\;\ref{resfig} are plotted as vertical lines (blue for spiral, red for bar). Solid is co-rotation, dotted the 4:1 and dot-dashed the 2:1 resonance. Models are shown at 400\,Myr of evolution, with an extra 350\,Myr time frame for the barred-spiral models.}
\label{SFradius}
\end{figure}

\begin{figure}
\includegraphics[trim = 0mm 5mm 0mm 5mm,width=80mm]{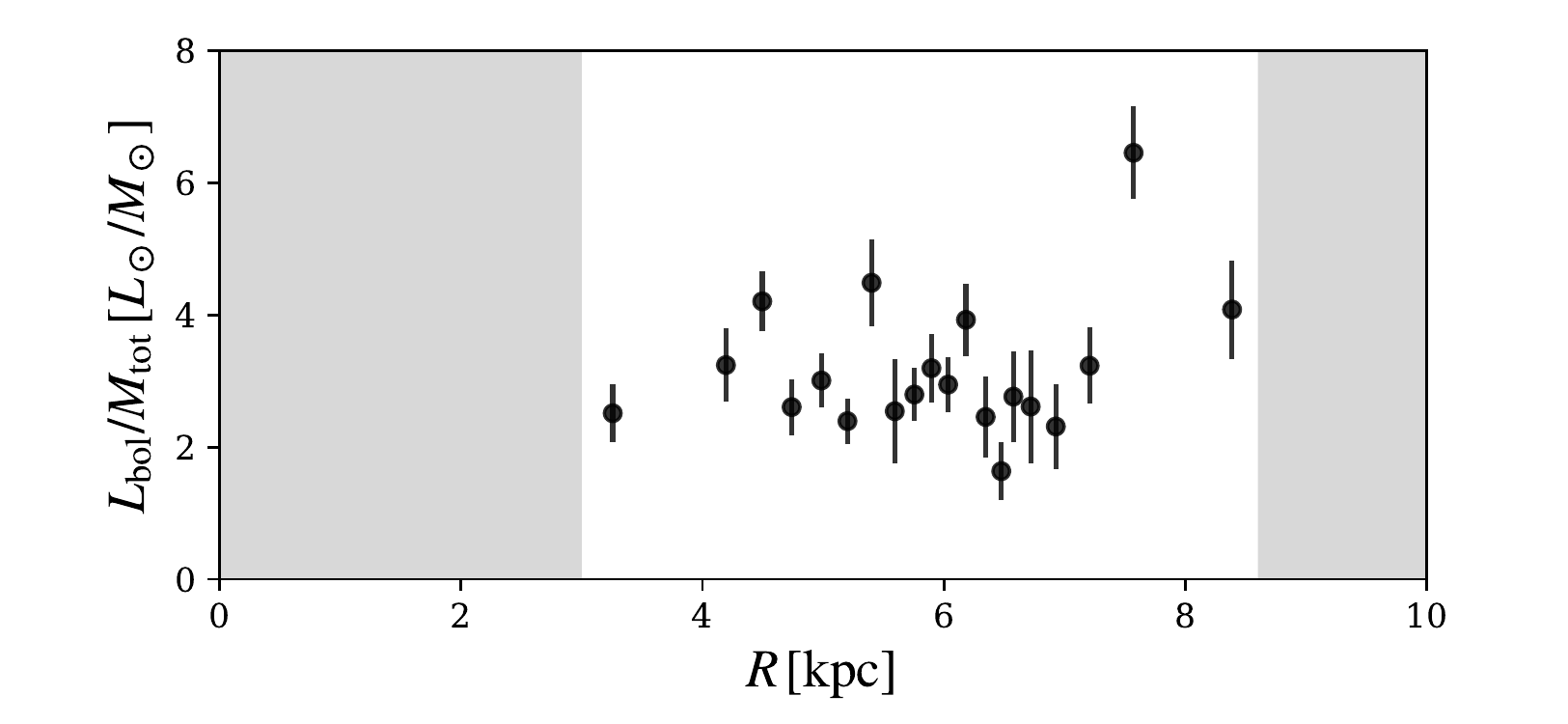}
 \caption{The same data as Fig.\;\ref{HigalLong} binned into bins of equal population in Galactocentric radius. The shaded region indicates the radial limits of the survey.}
\label{SFradiusHG}
\end{figure}

We collapse this data down further by looking at star formation as a function of Galactocentric radius in Figure\;\ref{SFradius}, calculated in annuli of 0.2\,kpc width. Each panel shows a different model, with the locations of the various resonances caused by the arms and bar as vertical lines. The barred models both exhibit clear peaks in star formation, loosely associated with the 2:1 and 1:1 resonances (ILR and CR in particular), while 4:1 features are nearly non-existent. The particularly strong peak just beyond the OLR is where the bar-driven arms wind up completely into a circular pattern (see Fig.\,\ref{ImsAll_G}). BrStr displays similar but stronger features, shifted outwards in radius due to the slower pattern speed. The Sp2 and Sp4 models have their CR and OLR resonances beyond the simulated domain, but the inner 2:1 and 4:1 resonances show mild increased star formation activity (clearer in Sp2 than Sp4). The ILR in particular does seem to correspond to a broad maxima (2:1 dot-dashed line in Sp2, and 4:1 dotted line in Sp4). The BrSp models are shown at two different time frames, as the bar-arm phase offset is time dependent. The BrSp4 model shows peaks in star formation at radii similar to a simple superposition of Sp4 and Br, perhaps not surprising as the ILR of the arms lies beyond the bar OLR. BrSp2 shows an interesting differences between time frames, seen clearest around 3\,kpc. At  400\,Myr the arms and bar are quite distinct (similar to the maps in Figs.\,\ref{ImsAll_G} and \ref{MySFGAL}), while at 350\,Myr the bar-driven arms and spiral potential arms are coincident, with the only spiral arms appearing to join to the bar end. This creates a strong density contrast in the gas compared to the interarm region, spurring on additional star formation. This effect may also be enhanced by the overlapping resonances where the arm ILR (2:1) lies close to both the CR and 4:1 resonance of the bar.

The distribution of star formation with Galactocentric radius for the BrSp models in Fig.\;\ref{SFradius} shows small scale variations in star formation rate in the mid-disc, not dissimilar to data from observations. Distances to Galactic star forming sources are generally hard to find, though measurements for $\Sigma_{\rm SFR}$ \citep{2012ARA&A..50..531K} and the distribution of GMCs \citep{2010ApJ...723..492R} both show gradually decreasing trends with $R$, with no clear small scale variations due to the relatively large uncertainties (seen also in external galaxies: \citealt{2016A&A...590A..44G}). As a specific example, Figure\;\ref{SFradiusHG} shows the luminosity to mass ratio of compact Hi-GAL clumps data of binned by Galactocentric radius \citep[see also][]{2016MNRAS.462.3123R}. The data does exhibit small scale undulations in $4{\rm \,kpc}<R<8{\rm \,kpc}$ range, but they do not stray far outside the error margins, being equally consistent with a flat distribution. The enhanced $L_\mathrm{bol} / M_\mathrm{tot}$ around $R_\mathrm{GC} \sim 7-8$\,kpc appears genuine, corresponding mainly to the tangents of the outer spiral arm (Sagittarius and Centaurus).

\subsection{Gaia DR2 comparison}

The release of Gaia DR2 brings with it excellent prospects for comparing models to Galactic structure \citep{2016A&A...595A...1G,2018A&A...616A...1G}. Some of  of the most striking visual results are the maps of the velocity fields of the stars in polar coordinates, as first shown in \citet{2018A&A...616A..11G}. We reproduce similar maps in Figure \ref{GaiaV} from the Gaia DR2 dataset, with a sample selection defined in Appendix \ref{Appx1}. The left panel shows the median radial velocity; ${V}_R$, centred about 0\,\kms, and the right the azimuthal velocity, ${V}_\phi$, centred about the circular velocity at the Solar position ($\approx 228$\,\kms{}). In each panel there are clear departures from uniform circular motion, which can be interpreted as the effect of non-axisymmetric perturbations such as arms and bars.

We present similar velocity field maps for models Sp2, Sp4, Br, BrSp2, BrSp4 in Figures \ref{GaSp2}, \ref{GaSp4}, \ref{GaBr}, \ref{GaBrSp2} and \ref{GaBrSp4}, constructed after 400\,Myr of evolution. All stars in the simulation are used, but we remind the reader this only corresponds to the young stellar population due to the lack of a live stellar disc. Stars in the simulation are binned into a Cartesian grid of resolution 200\,pc, and the median of the velocity distribution in question is calculated. Minima of spiral potentials are shown as grey lines, and dashed lines show mid-point between arms in Sp2 models. Non-axisymmetric features clearly show up in each of the simulated discs, highlighting the locations of arm and bar features. There also seems to be a similar correlation between features seen in ${V}_\phi$ and ${V}_r$ space in each of the plots, with arm features appearing in both maps. Looking at the Sp2 plot as an example, there are clear relations between ridges in velocity space and the arm minima, consistent with the standard density-wave like model of steadily rotation spiral arms \citep{1964ApJ...140..646L,1969ApJ...158..123R}. That is: the minimum of $V_r$ is associated with the potential minimum, whereas $V_\phi$ experiences a strong turning point. These patterns are different to those seen in transient/dynamic spiral arm models (see \citealt{2016MNRAS.460.2472B} for a comparison between the two arm models, in particular their figure 4). 

{
\begin{figure}
\begin{centering}
\includegraphics[trim = 10mm 15mm -10mm 0mm,width=100mm]{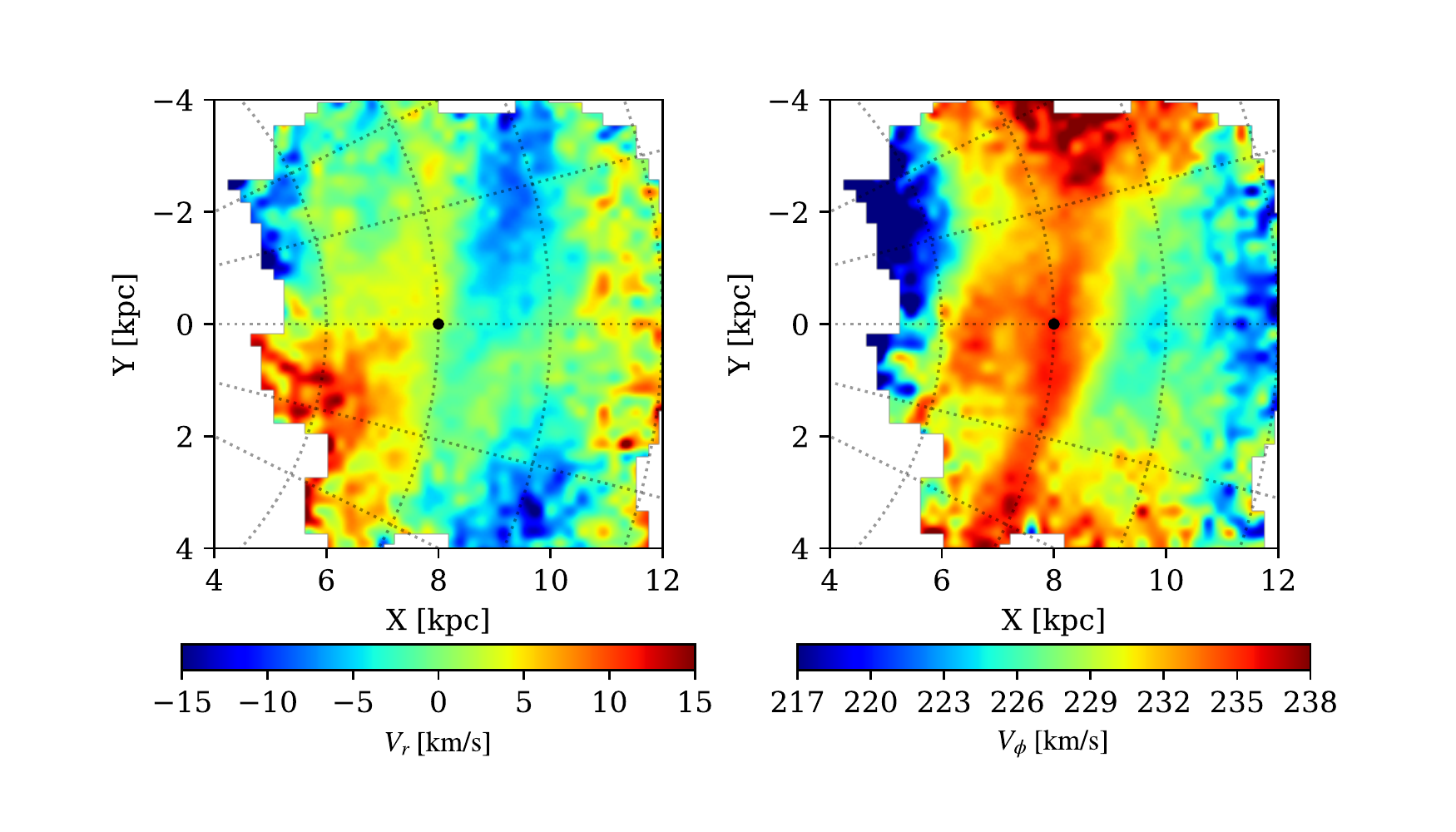}
 \caption{Median stellar radial velocity (left) and azimuthal velocity (right) fields from Gaia DR2. The sample selection is described in Appendix \ref{Appx1}.}
\label{GaiaV}
\includegraphics[trim = 10mm 15mm -10mm 0mm,width=100mm]{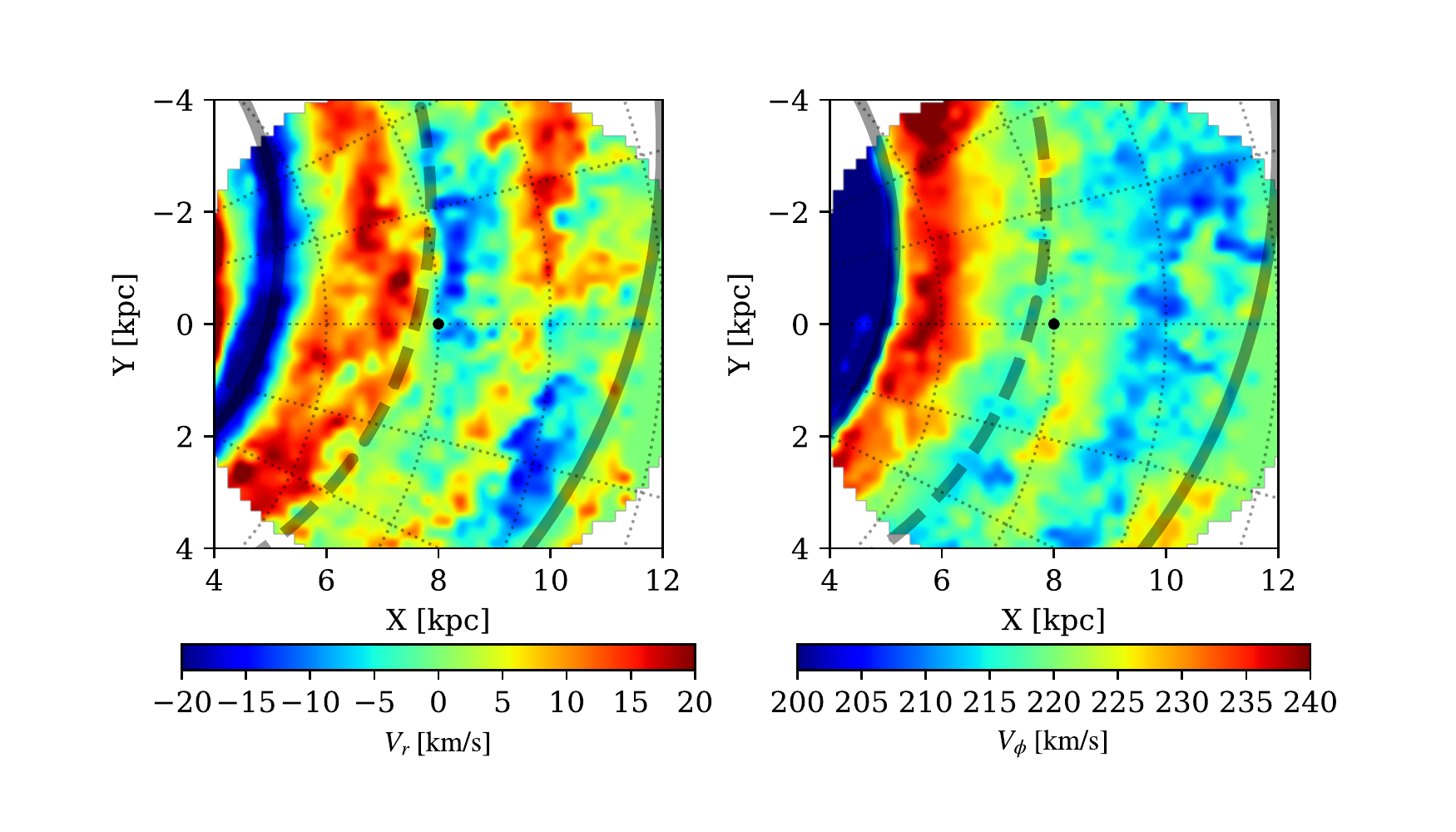}
 \caption{Stellar velocity maps for the Sp2 model after 400\,Myr of evolution. Solid line shows potential minimum, and dashed lines the mid-point between minima.}
\label{GaSp2}
\includegraphics[trim = 10mm 15mm -10mm 0mm,width=100mm]{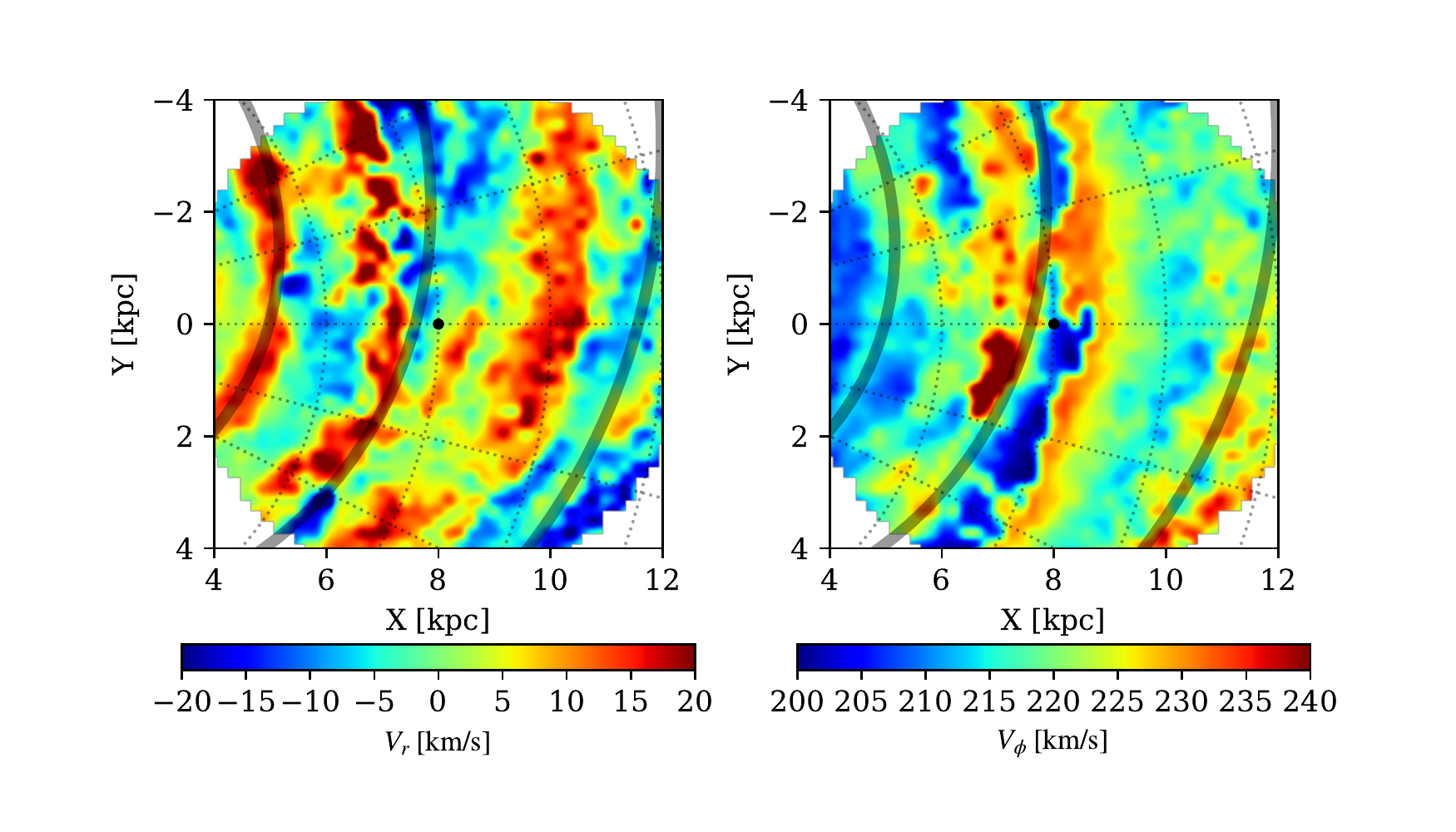}
 \caption{As Figure\;\ref{GaSp2} but for Sp4.}
\label{GaSp4}
\end{centering}
\end{figure}

\begin{figure}
\begin{centering}
\includegraphics[trim = 10mm 15mm -10mm 0mm,width=100mm]{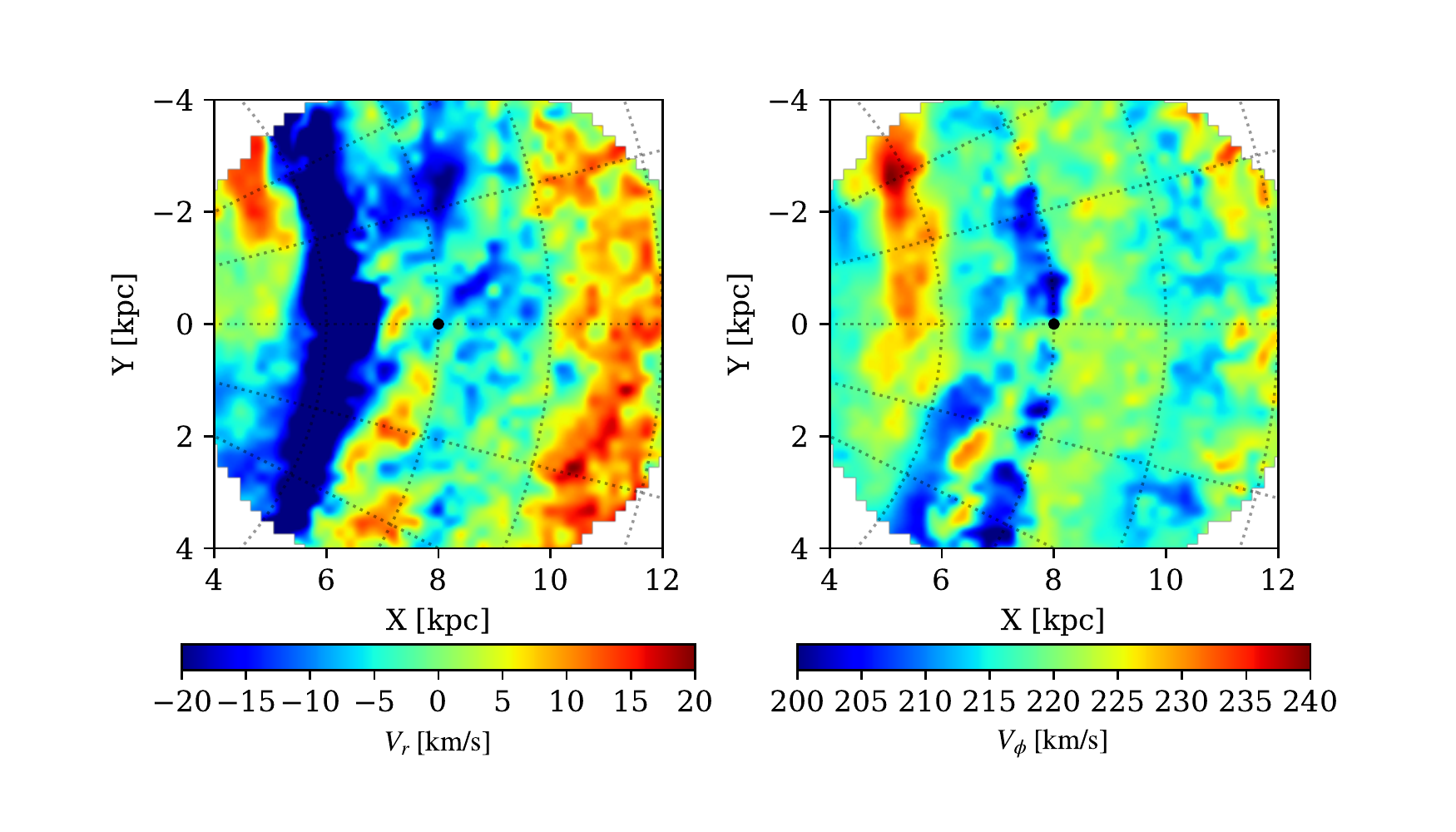}
 \caption{As Figure\;\ref{GaSp2} but for Br.}
\label{GaBr}
\includegraphics[trim = 10mm 15mm -10mm 0mm,width=100mm]{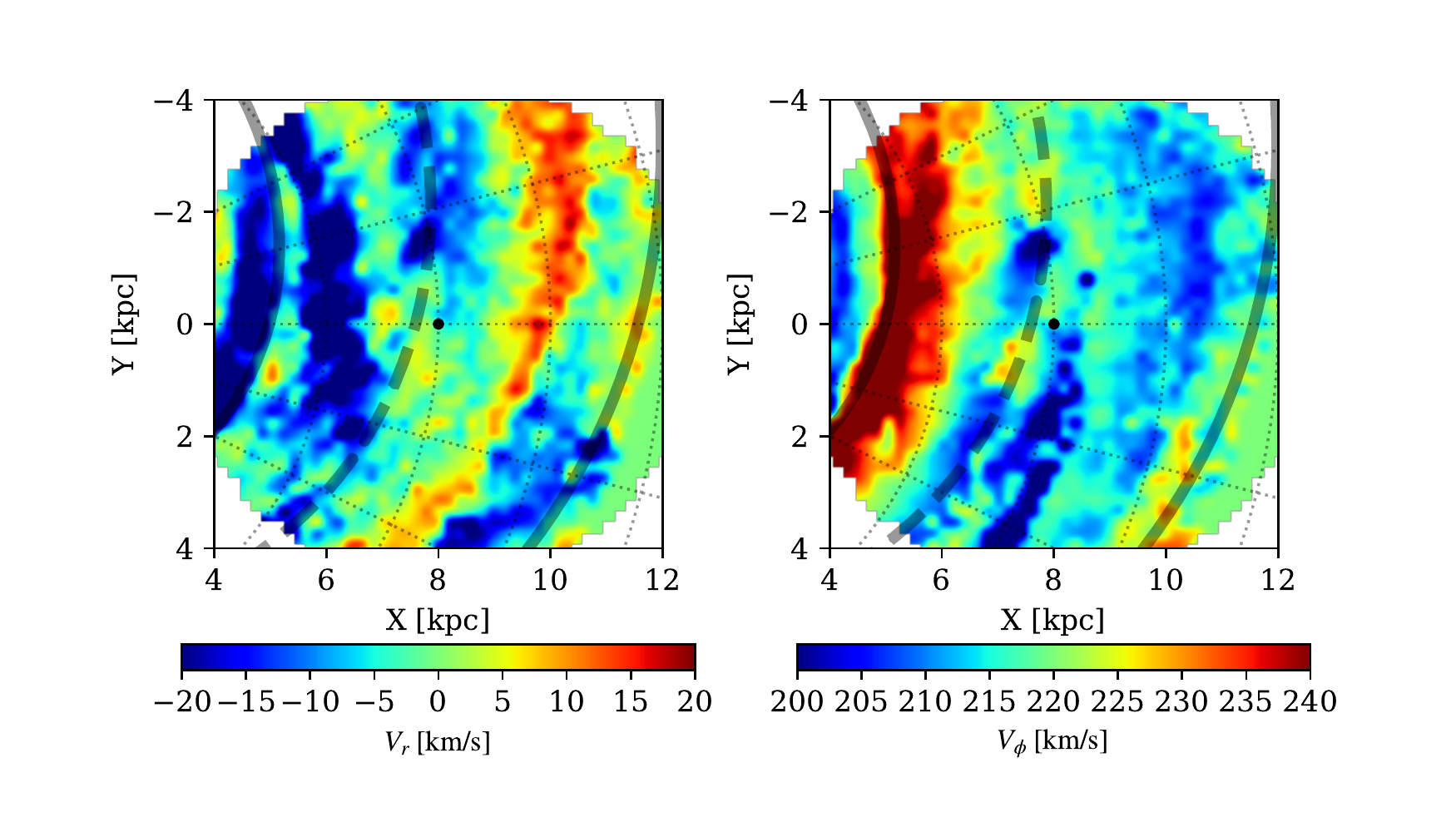}
 \caption{As Figure\;\ref{GaSp2} but for BrSp2.}
\label{GaBrSp2}
\includegraphics[trim = 10mm 15mm -10mm 0mm,width=100mm]{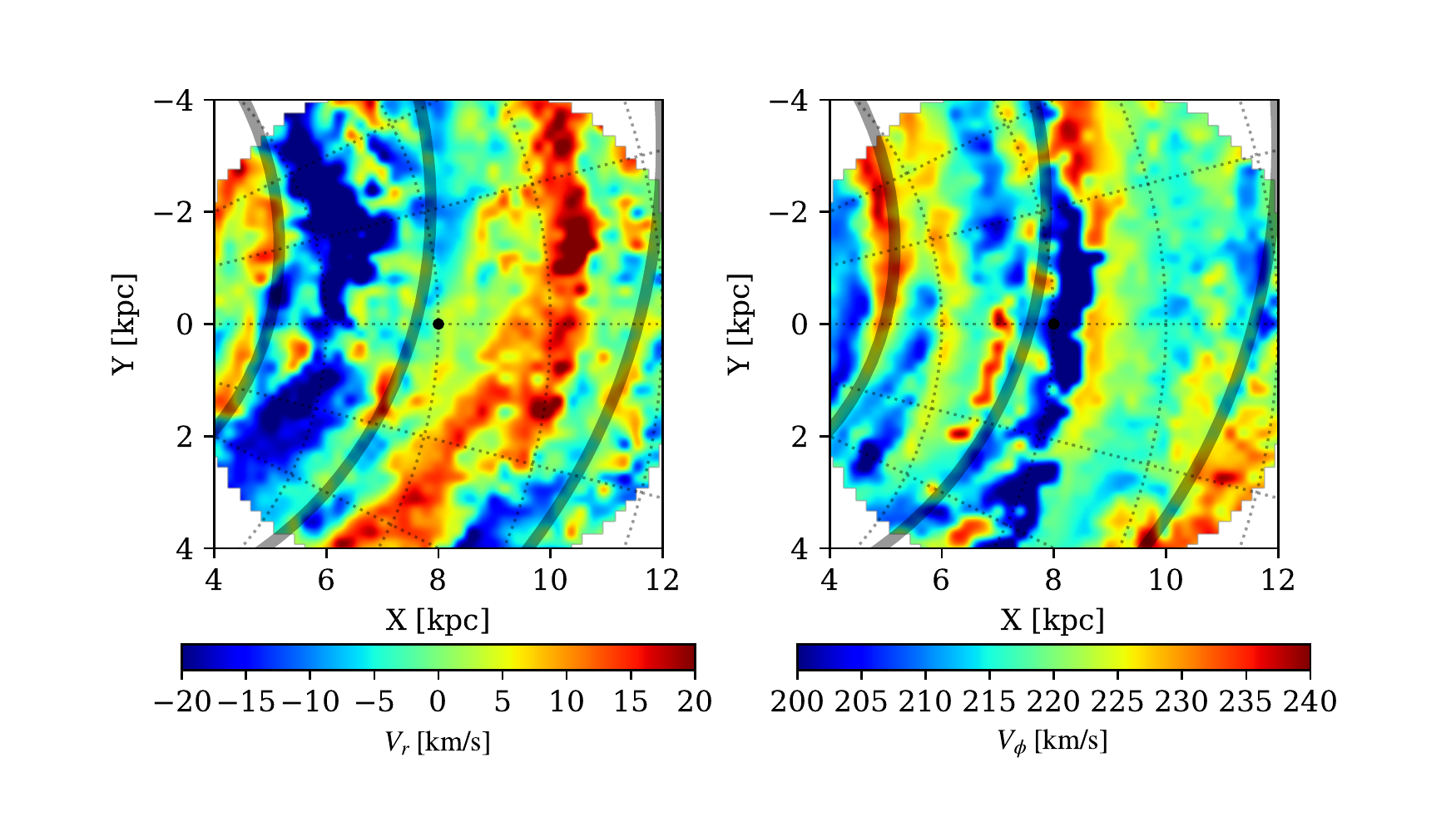}
 \caption{As Figure\;\ref{GaSp2} but for BrSp4.}
\label{GaBrSp4}
\end{centering}
\end{figure}
}

All the models have been rotated to the same orientation as Sp2, to allow for comparison between models. This orientation was chosen as the one where Sp2 was the closest match to the data, which also lines up the spiral minima with the model arms plotted in figure 19 of \citet{2018A&A...616A..11G} using the 2-armed model of \citet{2000A&A...358L..13D}. Maps apart from that of Sp2 serve to illustrate the differences between each model, and have not been specifically chosen to be a good match to the data. At this time the bar is orientated at 26\arcdeg{} to the Sun-Galactic centre line.

Arm features seen in the Sp4 model appear weakest out of all the models shown, with breadths of arms appearing considerably weaker (and less similar to the Gaia data). Top-down views of this model resemble a multi-armed or even flocculent disc such as M101, which gives credence to the picture that the Milky Way has a clear grand design structure with only a few, well defined, arm features. The addition of the rapidly moving bar (BrSp4) does alleviate this somewhat; allowing for the creation of stronger arm features in velocity space. The Sp2 model creates some clear velocity features of similar size to the Gaia data, with some time-frames in particular showing a very clear match to either the ${V}_\phi$ or ${V}_r$ data. However; reproducing both velocity fields proved difficult, as is evident from Figure \ref{GaSp2} where the azimuthal velocity is decently reproduced but the radial match is poorer. Though this may simply be due to some global factor such as pitch angle being incompatible with this particular arm.

\begin{figure}
\includegraphics[trim = 10mm 10mm 0mm 0mm,width=85mm]{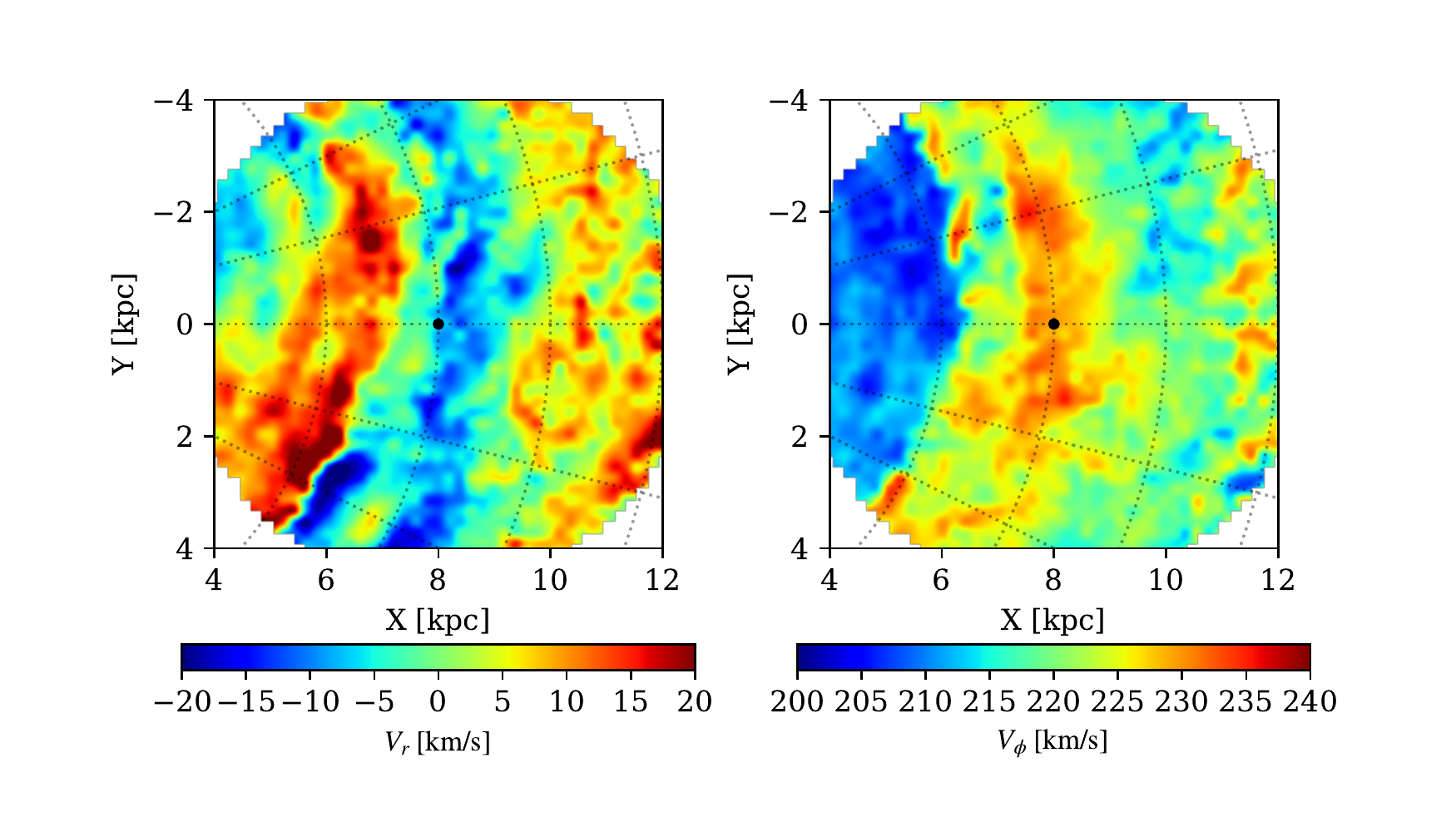}
\includegraphics[trim = 10mm 10mm 0mm 0mm,width=85mm]{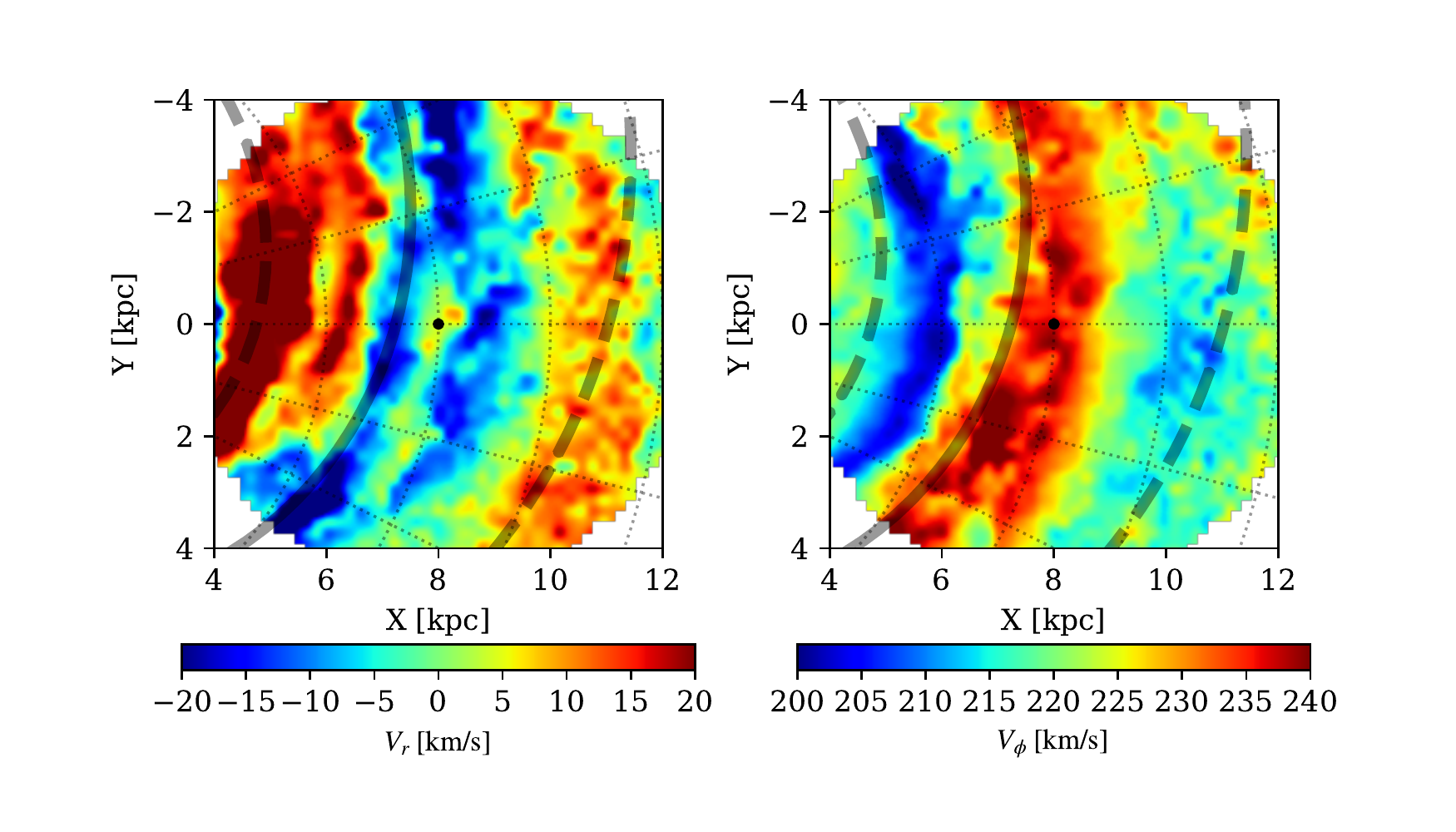}
\includegraphics[trim = 10mm 10mm 0mm 0mm,width=85mm]{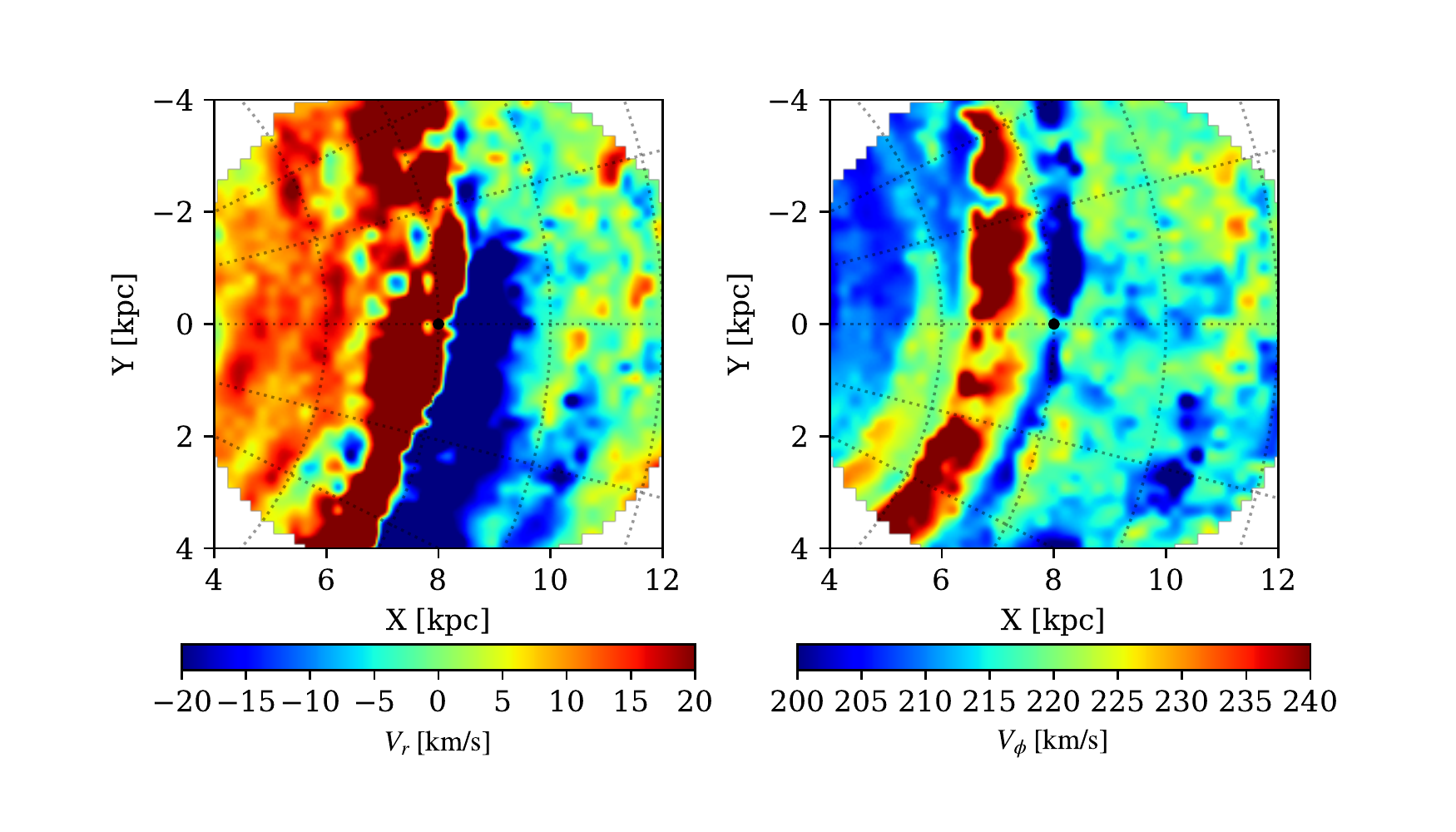}
 \caption{Additional velocity fields for the Br model (top) and BrSp2 (middle) but rotated by 90\arcdeg{} compared to Fig.\;\ref{GaBr}, giving a much better agreement with the Gaia DR2 maps. The map of BrStr that best appeared to represent the DR2 data is shown in the bottom panel.}
\label{GaBr_2}
\end{figure}

Adding a bar certainly alters the features, either creating weak inter-arm patterns or reinforcing the velocity perturbation around the arms. However, as the effect of the bar in this region is second order compared to the arms, this is not enough to match the patterns seen in the Gaia data. The bar model alone recreates reasonably good radial features in other time frames (see Fig\;\ref{GaBr_2}), but the azimuthal ones appear quite weak in comparison, indicating the need of some arm component to drive non-axisymmetric orbits in the mid/outer disc, with the gaseous response to the bar in this region simply not enough to perturb the velocity field. Note however that a slower bar model will result in a more radially extended gas response, which would drive clearer structures in the outer edges of the disc than seen here. The combination of bar and spiral features appears to be a simple superpostion of velocity fields. For instance, Sp2 clearly has the $X=7$\,kpc high radial velocity component reduced significantly by the Br contribution in the BrSp2 model, while the azimuthal component has been enhanced at the same location due to the Br and Sp2 models both having strong positive components. Recall the arms driven by the bar are not log-spiral in shape, so their relative $V_r$ and $V_\phi$ response is different to that of the arms.

We highlight to the reader that we are not insinuating that the barred-spiral paradigm is unfavourable to the data, merely that the interplay between these features can strongly impact the observed velocity distribution (at least in the younger stellar population). In order to discern the true validity of structure models a full parameter sweep through different arm and bar morphologies and evolution epochs would be needed as well as an inclusion of the older stars to consistently compare, which is outside the breadth of this study. To illustrate this point we shown additional maps of Br and BrSp2 in Figure\;\ref{GaBr_2} (top and middle panels), at a different orientation to that shown in Fig.\;\ref{GaBr} so that features better match the Gaia maps, rather than simply being selected at the orientation Sp2 gave a good match. The maps in this figure reproduce the broad features well, though  the bar is now angled angled at 110\arcdeg. It may simply be that the pattern speed adopted here is not conducive to exactly matching all of such features, at least in the young stars. The bottom panel shows map of BrStr, rotated to give the best agreement with the DR2 data. While the bar end still lies in the first quadrant (angled at 82\arcdeg), it creates features in $V_{\phi}$ and $V_{r}$ space that are seemingly incompatible with the Gaia data, with the azimuthal peak in velocity being closer to the Galactic centre than the radial one. This difference compared to Br is due to the slower pattern speed and resulting different radial response of the disc, so that the solar position lies very much deep within the bar driven arm, while for Br it is closer to the region where the arms are wrapping up (this is also why features appear narrower in velocity compared to those in Br).

There are quite large ranges in the binned velocities compared to the observed data. This over-heating of young stars can be due to a number of reasons. The spiral potential used may be too strong, introducing too large streaming motions in $R$ and $\phi$. Additionally; just picking up on the young stars mean their velocities track the gas (due to the nature of the star particles spawning from gas particles directly), which has significantly more turbulent motion than the stars, despite being kinematically colder than using an older stellar population. i.e. velocities are inherited from spiral shocks around the arm, and are influenced by nearby supernova events and cloud-cloud interactions. Similar maps were made from the live stellar disc simulations from \citet{2017MNRAS.468.4189P}, which showed a much lower variance in the mean velocities when including an inherent ``old" stellar population, more in line with Gaia data. Similarly, the lack of an old population meant that attempts to measure velocity dispersions suffered from low counts in some outer bins, and would thus require a simulation with a resolved stellar component across the entire age spectrum to properly compare. We attempted to create similar maps of the observed stellar distribution only using a young stellar population, but sources with known ages were too few to garner meaningful velocity maps.

A number of recent studies have investigated the various influences on the velocity distribution in Gaia-sized fields with numerical simulations. \citet{2015MNRAS.453.1867G} performed a numerical study of numerous different spiral arm models with live $N$-body discs. They find that velocity fields can be highly sensitive to the nature of spiral arms, be they bar driven, density wave-like, or seeded by disc instabilities. In \citet{2019MNRAS.485.3134L} the authors use $N$-body simulations of a Milky Way type disc and the interaction with the in-falling Sagittarius-like dwarf galaxy, producing a remarkable correspondence to the Gaia DR2 velocity maps (see also \citealt{2019MNRAS.486.1167B}). Though the maximum velocity variations are nearly double what is seen in the DR2 data, similar to our maps.

Preliminary efforts were made to explore the young stellar kinematics in greater detail, such as examining features in the UV velocity plane \citep{1998AJ....115.2384D,2010LNEA....4...13B} and orientation of the stellar velocity ellipsoid \citep{1998MNRAS.298..387D,2012ApJ...746..181S}. However, our gas resolution and limitation to the young stellar population posed too strong restrictions on number of stars in the solar neighbourhood for reliable measurements, especially exterior to the Solar radius. Such features will be the subject of a future study.

\section{Analytic model comparison}
\label{sec:model}
\begin{figure}
\includegraphics[trim = 10mm 10mm 0mm 0mm,width=85mm]{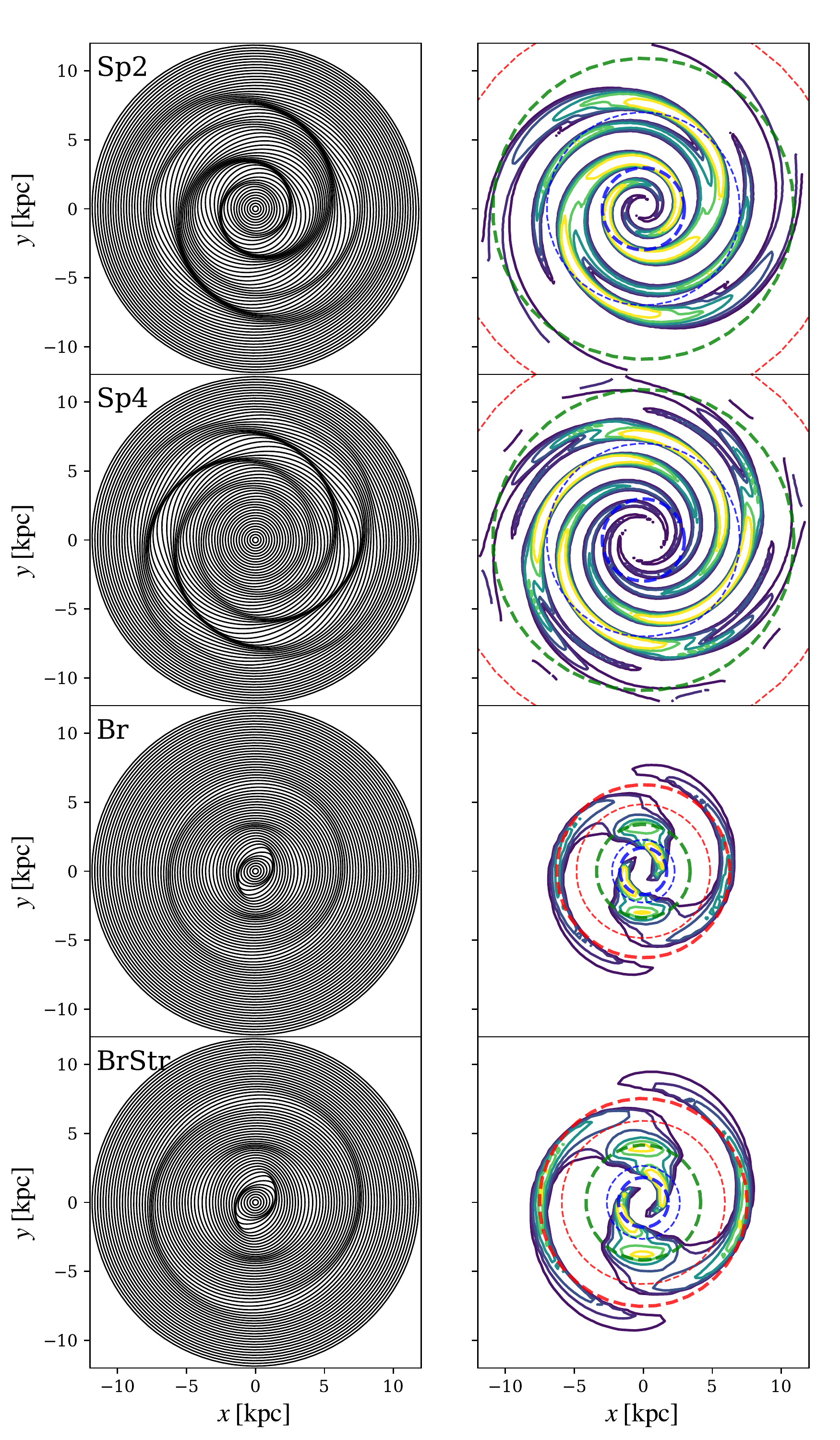}
 \caption{ A selection of orbital families (left) and the resulting density perturbation (right) for the four different non-axisymmetric potentials considered in this work. The green dashed circle denotes CR. Blue inner and outer circles are the inner 2:1 and 4:1 resonances respectively, and the red inner and outer circles are the outer 4:1 and 2:1 resonances respectively. The bars are orientated vertically.}
\label{AnalyticOrbs1}
\end{figure}

We can use simple analytic models to illustrate the effect of bars and spiral on the ISM, as shown in this study. Such models have been presented in \citet{1994ASPC...66...29L} and \citet{1994PASJ...46..165W}, where the authors use the epicyclic approximation to study orbits of packets of gas subject some non-axisymmetric background perturbation. The dissipative nature of the gas is included via a damping/friction term in the equations of motion, which perturb the motion of gas particles in the frame of the perturbation by $r=r_0+\xi$ and $\theta=\theta_0+(\Omega+\Omega_p)t+\eta/r_0$, where $r_0$ and $\theta_0$ are the undamped orbital solutions. While the magnitude of the damping term (i.e. damping frequency, $\lambda$) is a somewhat free parameter, it is often chosen to be a number close to the pattern speed. These approaches suffer from a singularity at CR, which was addressed by \citet{2012MNRAS.421.1089P} (hereafter PF2012) through the inclusion of an additional softening factor. Such models have been used in the past to constrain bar properties in external galaxies \citep{1999ApJS..124..403S,2014MNRAS.438..971P} as well as the bar and spiral features of the Milky Way \citep{2015MNRAS.451.3437S,2015PASJ...67...69S,2015PASJ...67...70H}.

\begin{figure*}
\includegraphics[trim = 10mm 0mm 0mm 0mm,width=140mm]{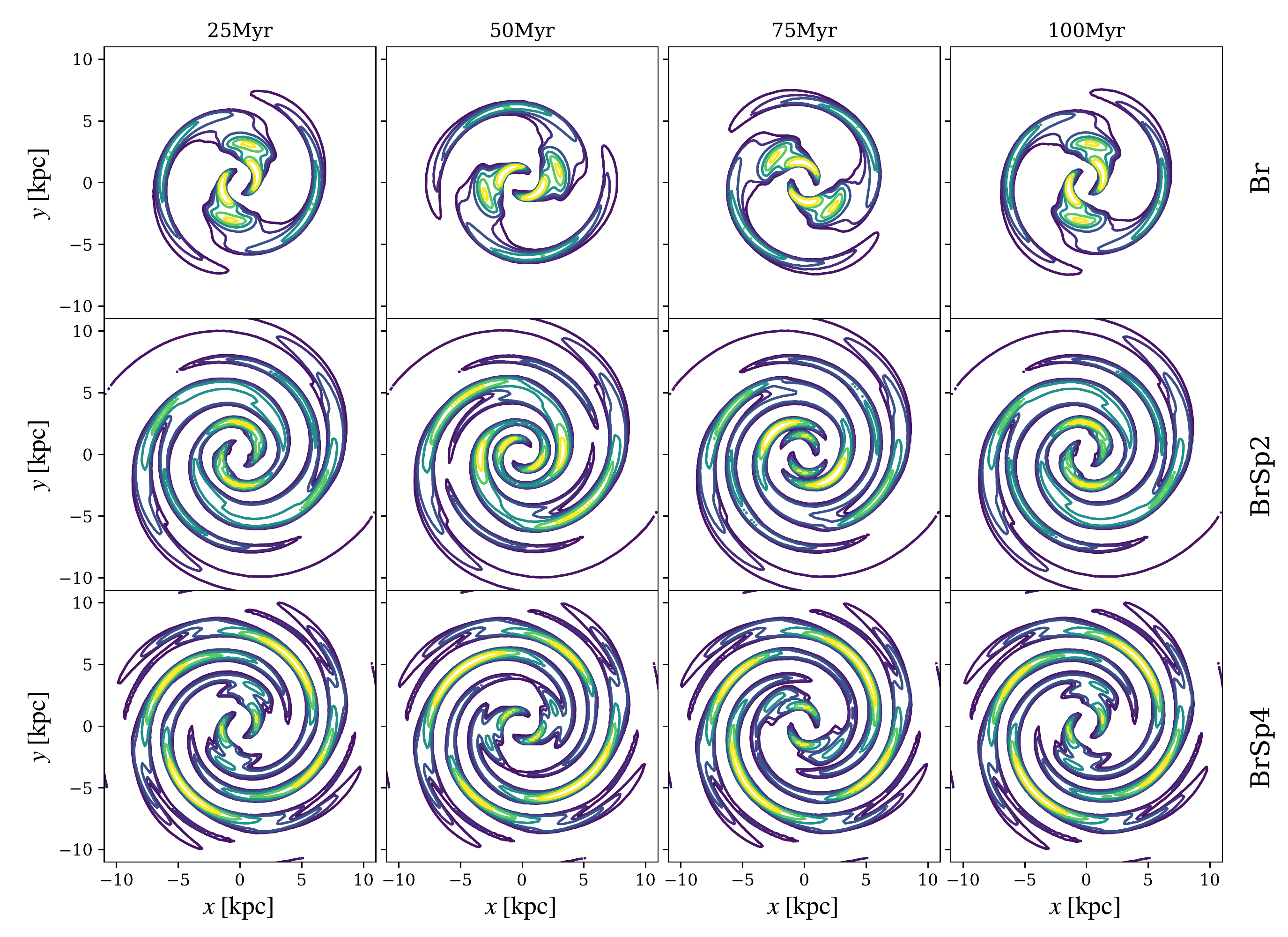}
 \caption{ Analytic model of the perturbation from the bar added to that of the different spiral models. Each row indicates a different model, and each column a different time, which in turn determines the arm-bar phase offset.}
\label{AnalyticOrbs2}
\end{figure*}

We follow the approach of PF2012 altered to include our specific axisymmetric potentials, bars and spiral arm features. 
The dispersion is chosen to scale with the epicycle frequency, $\lambda = \Lambda \kappa$, by a factor $\Lambda=0.1$, following \citet{1994PASJ...46..165W,1999ApJS..124..403S}, rather than in PF2012 where the authors use a seemingly arbitrary linear scaling law or constant value throughout the disc. The CR softening frequency, $\epsilon$, is a free parameter that we take to be $\kappa/5$ for the bars and $\kappa/2$ for the arms, values that appeared to reproduce the structures in the simulations while avoiding orbital crossings. The density perturbation is calculated via the continuity equation, as in PF2012. While the quantitive results will change for different choices of $\Lambda$ and $\epsilon$, the morphological features are rather robust to changes in these values.

We show the orbital families and over-densities for our Sp2, Sp4, Br and BrStr potentials in Figure\;\ref{AnalyticOrbs1}. Several of the resonances are over-plotted as coloured circles. The resemblance between these maps and the gas density in Fig.\;\ref{ImsAll_G} is quite remarkable. Sp2 and Sp4 have radial extent of the gas over-density near identical to that of the simulation, with Sp2 extending to smaller radii than Sp4. Branches are clearly present in the Sp2 density maps, a feature that has been in several past simulations, including \citet{2003ApJ...596..220C,2004MNRAS.350L..47M,2014arXiv1406.4150P}, each implementing a different spiral potential model. These branches are faintly present in our simulations, appearing clearest in the young stars (Figs. \ref{sf_events} and \ref{MySFGAL}). The response to the bar is similar to what has been shown in other studies, \citep{1994PASJ...46..165W,1994ASPC...66...29L}, with the steady phase shift in gas orbits between the ILR and OLR manifesting as a kind of spiral pattern. With the slower, stronger bar the resonances extend to larger radii, with both bar maps showing excellent agreement with the simulated gas density.

As a basic illustration of the time-dependent interplay between these features, we combine the density perturbations shown in Fig.\;\ref{AnalyticOrbs1}, with the appropriate time-dependent phase offset, in Figure\;\ref{AnalyticOrbs2}. Maps of the combined density perturbation are shown in the frame of the spiral pattern. The bar rotates faster than the spiral, as seen in the time series in the top row. The second and third rows show the BrSp2 and BrSp4 combinations, respectively. The impact of the different pattern speeds is clearly seen by comparing features between time-frames. As the bar rotates in BrSp2 it enhances either the primary spiral pattern (as seen at 50Myr) or enhances the inter-arm branches of the 2-armed spiral, creating a 4-armed morphology. This agrees with what is seen in the simulations in this study. For BrSp4 the bar pattern creates an alternation in the strength of a pair of the arms, also similar to what is seen in the simulations. This is clearest in the lower left quadrant, with the arm around $(-5,-5)$kpc changing in strength as the bar passes by.

\section{Conclusions}
\label{sec:conc}
We have performed simulations of gas embedded in background potentials representing possible morphologies of the Milky Way galaxy, with the aim of investigating the influence of arm and bar feature on the star forming ISM. Spiral arms and bars are clearly dominating features in moulding the star forming ISM, with young stellar material strongly associated with the arm and bar features. The star forming regions are not uniformly coincident with the bottom of the spiral arms, showing offsets that vary as a function of radius, similar to what has been seen in previous works that have studied the location of shocked gas with respect to arms (e.g. \citealt{2004MNRAS.349..909G,2015PASJ...67L...4B}). The inclusion of an inner bar moving at a higher pattern speed than the arms (as is expected in the Milky Way) acts to disrupt the locations of star formation from what is seen using just arm potentials. This is especially prominent for the two-armed models, where there is significant arm/bar resonance overlap. The faster bar can have a direct impact on the secondary features, such as locations on inter-arm spurs and arm branches. There is a mild beat-like frequency observed in the star formation due to the overlap between the bars and arms in both two- and four-armed models. These could explain the dichotomy between major and minor spiral arms seen in different tracers in the Milky May discussed in the literature \citep{2001ApJ...556..181D,2008ASPC..387..375B}, where the major arms are observed at an epoch where the spiral arm and bar-driven features overlap. Such patterns would not be expected in short lived transient arm features, as individual spiral patterns would wind up on timescales similar to the bar rotation (e.g. \citealt{2012MNRAS.426..167G}).

Top-down maps of the locations of star formation show a variety of features in the inter-arm regions, as well as a lack of 1:1 tracing of arms in a few instances, with the morphology again being a function of the position of the bar relative to the arms due to the differing pattern speeds. Comparison to several observational analogues shows similar features, such as a lack of star formation in the Norma arm in the fourth quadrant. Spiral arm tangencies are well traced by the young stars in the outer disc, though the inner-disc is somewhat more ambiguous, especially in the four-arm models. The bar can act to strengthen arm features in the inner disc, making some tangencies more pronounced (e.g. $l=20^\circ$ tangent in Sp2 and BrSp2), though also diminishing the strength of others depending on the exact orientation of the bar with respect to the arms.

Analogues to the recent Gaia DR2 velocity maps were constructed, though with the caveat of being confined to the young stellar population. Spiral arms clearly show up in both azimuthal and tangential velocities, and once again the bar can act to alter the features considerably. While we do not attempt to fit to the DR2 maps or perform a parameter sweep, the simulated maps show promising reproductions of features seen in the data for both barred and unbarred discs. The four-armed models show quite a poor match to the data, with features that are too thin compared to the observed maps, and too many velocity minima in the Gaia field. The Gaia data clearly holds excellent potential for constraining the arm and bar morphology, highlighting the continued need for simulations, especially those that can additionally constrain the gas distribution and star forming properties of the Galaxy.

\section*{Acknowledgments}
The \textsc{python} package \textsc{pynbody} \citet{pynbody} was used for density renders of the SPH particle data, and \textsc{numpy} and \textsc{scipy} for the analysis. We thank \citet{2017MNRAS.471.2357W} for the public release of the SPH code, \textsc{gasoline2}, and the authors for their advise in running the calculations shown here.

We thank the anonymous referee for their comments and suggestions that greatly improved this manuscript. Many thanks to Anthony Brown for help related to using the Gaia DR2 dataset. We thank T. Naylor, J. Vickers, K. Wada and A. Quillen for helpful comments and discussion regarding this work. MCS acknowledges financial support from the National Key Basic Research and Development Program of China (No. 2018YFA0404501) and NSFC grant 11673083.

This project was developed in part at the 2019 Santa Barbara Gaia Sprint, hosted by the Kavli Institute for Theoretical Physics at the University of California, Santa Barbara. This research was supported in part at KITP by the Heising-Simons Foundation and the National Science Foundation under Grant No. NSF PHY-1748958. 

This work has made use of data from the European Space Agency (ESA) mission
{\it Gaia} (\url{https://www.cosmos.esa.int/gaia}), processed by the {\it Gaia}
Data Processing and Analysis Consortium (DPAC,
\url{https://www.cosmos.esa.int/web/gaia/dpac/consortium}). Funding for the DPAC
has been provided by national institutions, in particular the institutions
participating in the {\it Gaia} Multilateral Agreement.

This work has made use of the Herschel/Hi-GAL compact source catalogue, VIALACTEA - The Milky Way as a Star Formation Engine, reference VL-IAPS-DI-2016-004, \citep{2016A&A...591A.149M}.

\bibliographystyle{mnras}
\bibliography{Pettitt_MWstars.bbl}



\appendix
\section[]{Gaia DR2 selection}
\label{Appx1}

We select stars with full 6D velocity information, vetting by spurious velocities as recommended in \citet{2019MNRAS.486.2618B}. Sources are restricted to those with relative parallax errors better than 20\%, must have at least five transits to compute the radial velocity, and exclude sources with bright neighbours within  6.4  arcseconds (70365 sources). Sources with Renormalised Unit Weight Error (RUWE) greater than 1.4 were discarded, and magnitudes were corrected following \citet{2018A&A...616A...4E} and \citet{2018A&A...619A.180M}. A giant subsample was selected to highlight large scale structure, using the same colour-magnitude selection as used in \citet{2019MNRAS.485.3134L}. The in-plane velocity maps were then made by selecting sources within $\pm 200$pc of the midplane, resulting in a sample of 861259 sources. The Astropy package was used to convert to a galacto-centric coordinate system \citet{2018AJ....156..123A,2013A&A...558A..33A}. This sample selection and vetting was heavily based on the `gold sample' created by A. Brown at the 2019 Santa Barabra Gaia Sprint: \url{https://github.com/agabrown/gaiadr2-6dgold-example}.

The resulting sample successfully re-creates familiar plots from the Gaia DR2 literature, and produces top-down velocity maps with near identical to those of \citet{2018A&A...616A..11G} and \citet{2019MNRAS.485.3134L}, despite the latter using the improved parallaxes of \citet{2018AJ....156...58B}.

We additionally attempted to construct maps using only a younger stellar population of DR2 stars, defined by applying a colour cut-off of $G_{\rm BP}-G_{\rm RP}<0.5$ to pull out only OBA stars (as indicated in \citealt{2018A&A...616A..10G}). This only returned 21454 sources, before any latitude vetting, and most of which were confined to 1\,kpc from the Solar position, making this subselection unsuitable for the task of comparing large-scale galactic structure.

\section[]{Arms and a stronger bar}
\label{Appx2}

\begin{figure}
\includegraphics[trim = 10mm 10mm 0mm 0mm,width=85mm]{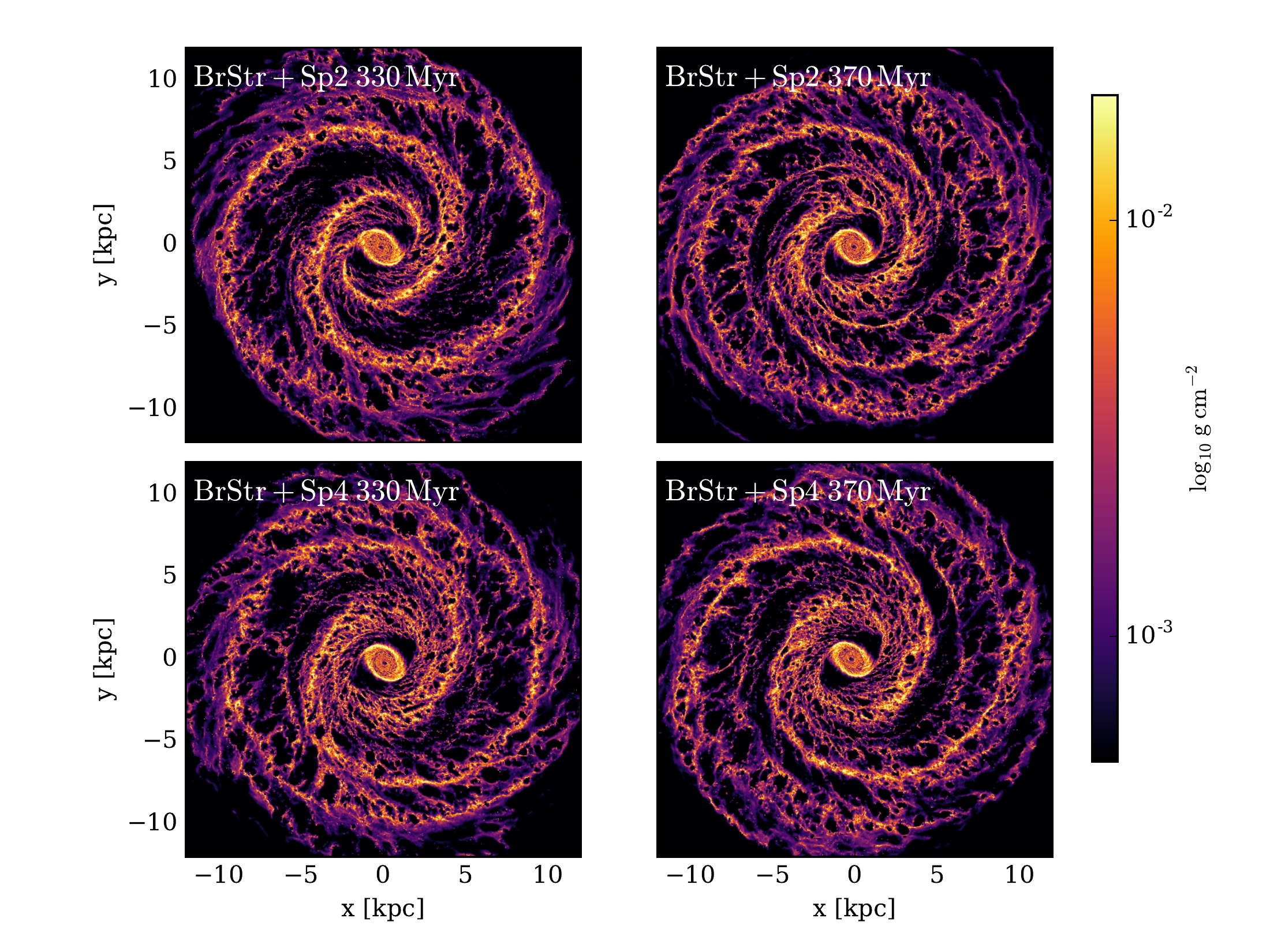}
 \caption{ The gas response to the combination of the strong bar model (BrStr) with the two and four spiral models (top: Sp2, bottom: Sp4). Two different time-frames are shown (left: 330\,Myr, right: 370\,Myr), illustrating the time-dependence on the resulting morphology.}
\label{StrongBar}
\end{figure}

\begin{figure}
\includegraphics[trim = 10mm 10mm 0mm 0mm,width=85mm]{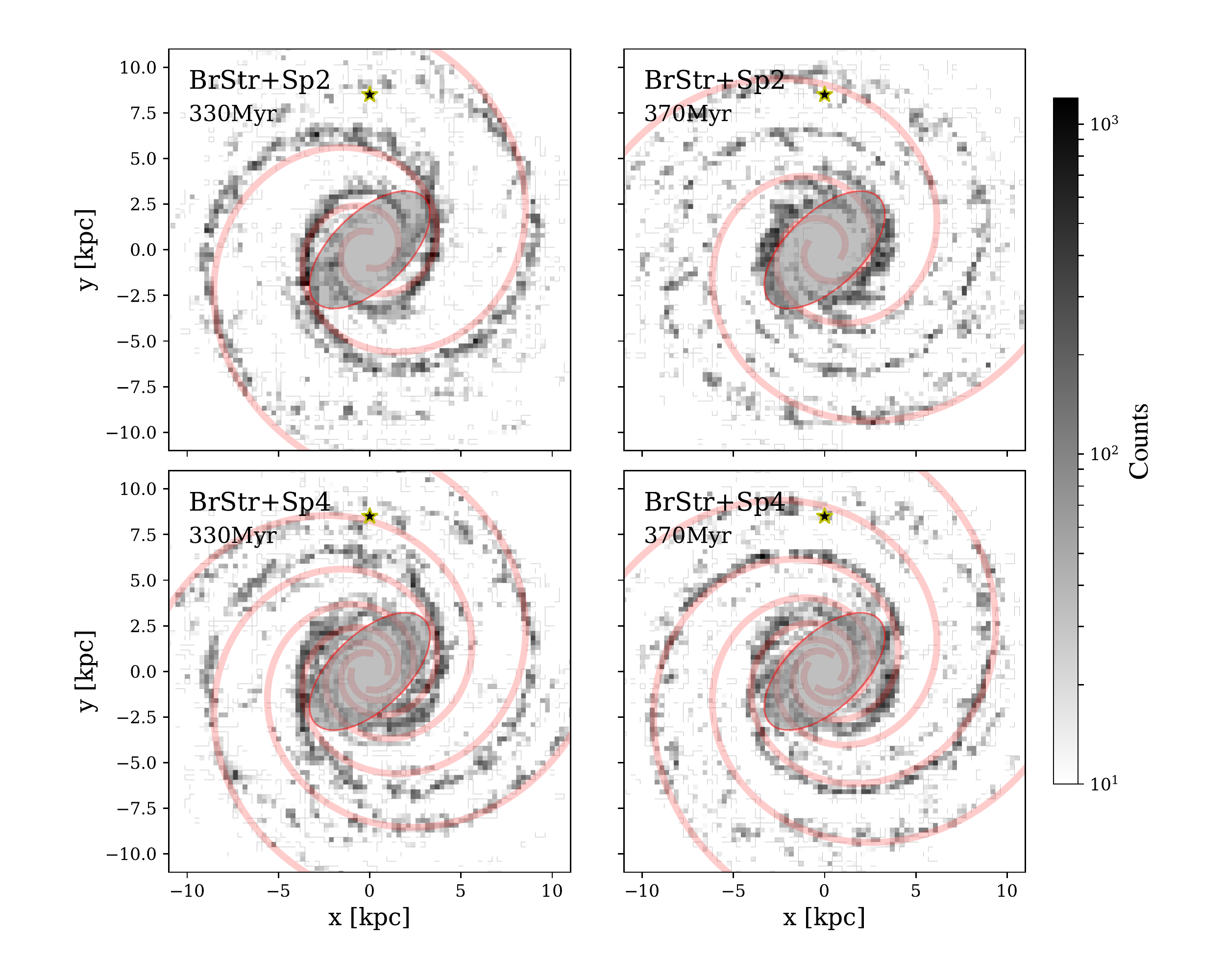}
 \caption{ Locations of star formation for BrStr+Sp2 and BrStr+Sp4, analogous to those of Figure\;\ref{MySFGAL} in the main text, at 330\,Myr and 370\,Myr. Red lines and grey oval indicate the orientation of the bar and spiral arm potentials, where the bar has been orientated at 45\arcdeg{} to the Sun-Galactic centre line.}
\label{StrongBarSF}
\end{figure}

Additional calculations with the BrStr and Sp potentials were performed. Figure\;\ref{StrongBar} shows the resulting gas morphology for both BrStr+Sp2 and BrStr+Sp4 at two different epochs to highlight the time-dependence of features, with star formation locations shown in Figure\;\ref{StrongBarSF}. The strong bar clearly has a greater influence on the structure of the ISM, relegating the spiral-driven arms to secondary features. The bar-driven arms clearly show up in the star formation map, with only a weak correlation between star formation events and the spiral potential minimum when the arms and bars are not aligned.

\begin{figure*}
\includegraphics[trim = 0mm 0mm 0mm 0mm,width=120mm]{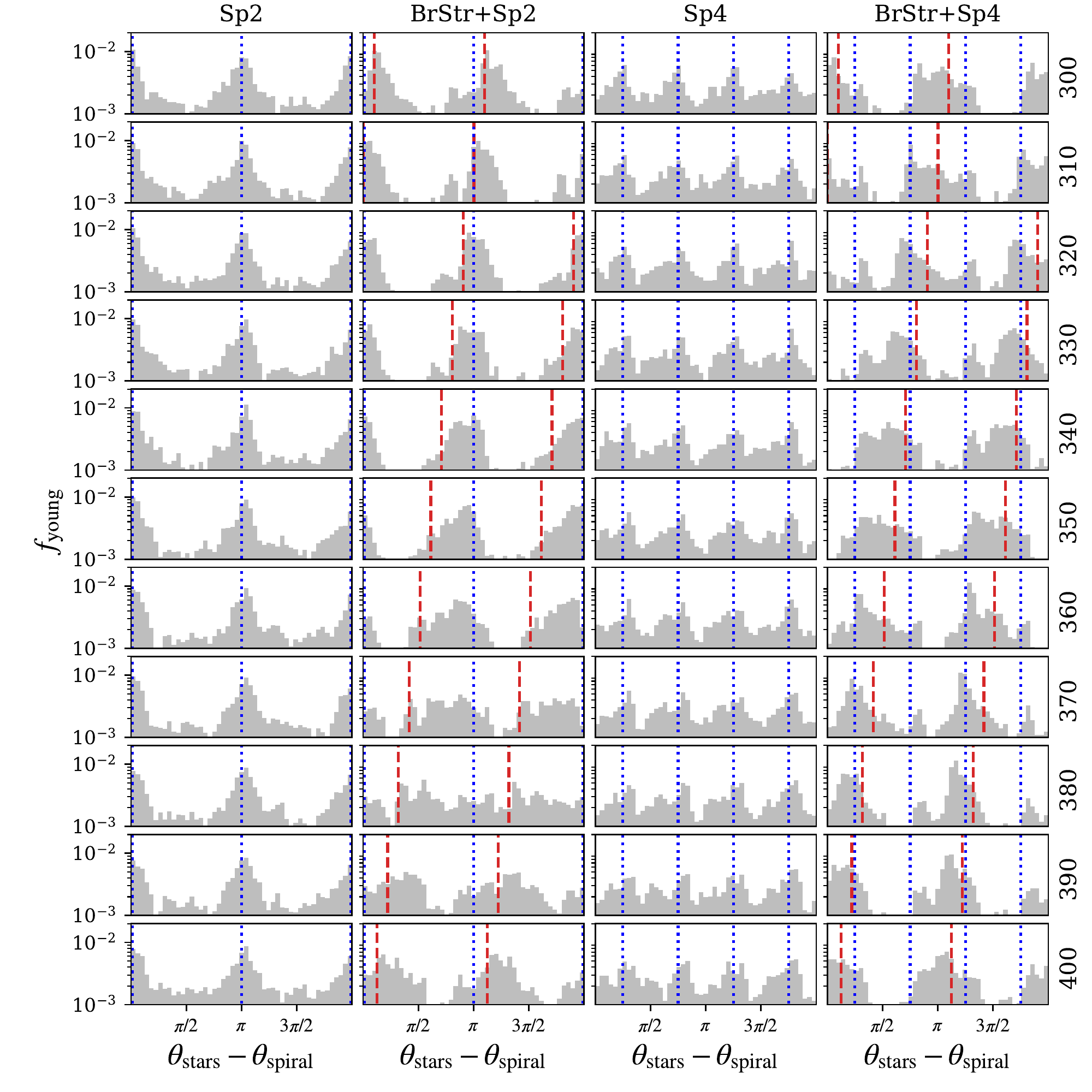}
 \caption{
Locations of young stars in the simulations with a stronger bar model combined with the spiral potentials. Young stars (less than 50\,Myr old) at radii $>2.5$\,kpc are placed in azimuthal bins phase-shifted such that the $x$-axis is the offset with respect to the arm potential. Each row indicates a different time in the simulation (300--400\,Myr). The dashed blue vertical lines indicate the arm potential, and the red lines the bar potential which moves to the left in the rest-frame of the arms due to faster rotation speed. Note that the red dashed line indicating the orientation of the bar takes longer to span a full $2\pi$ radians compared to that in Figure\;\ref{SFbeats} in the main text due to the lower pattern speed. }
\label{BeatsStrong}
\end{figure*}

\begin{figure}
\includegraphics[trim = 10mm 10mm 0mm 10mm,width=85mm]{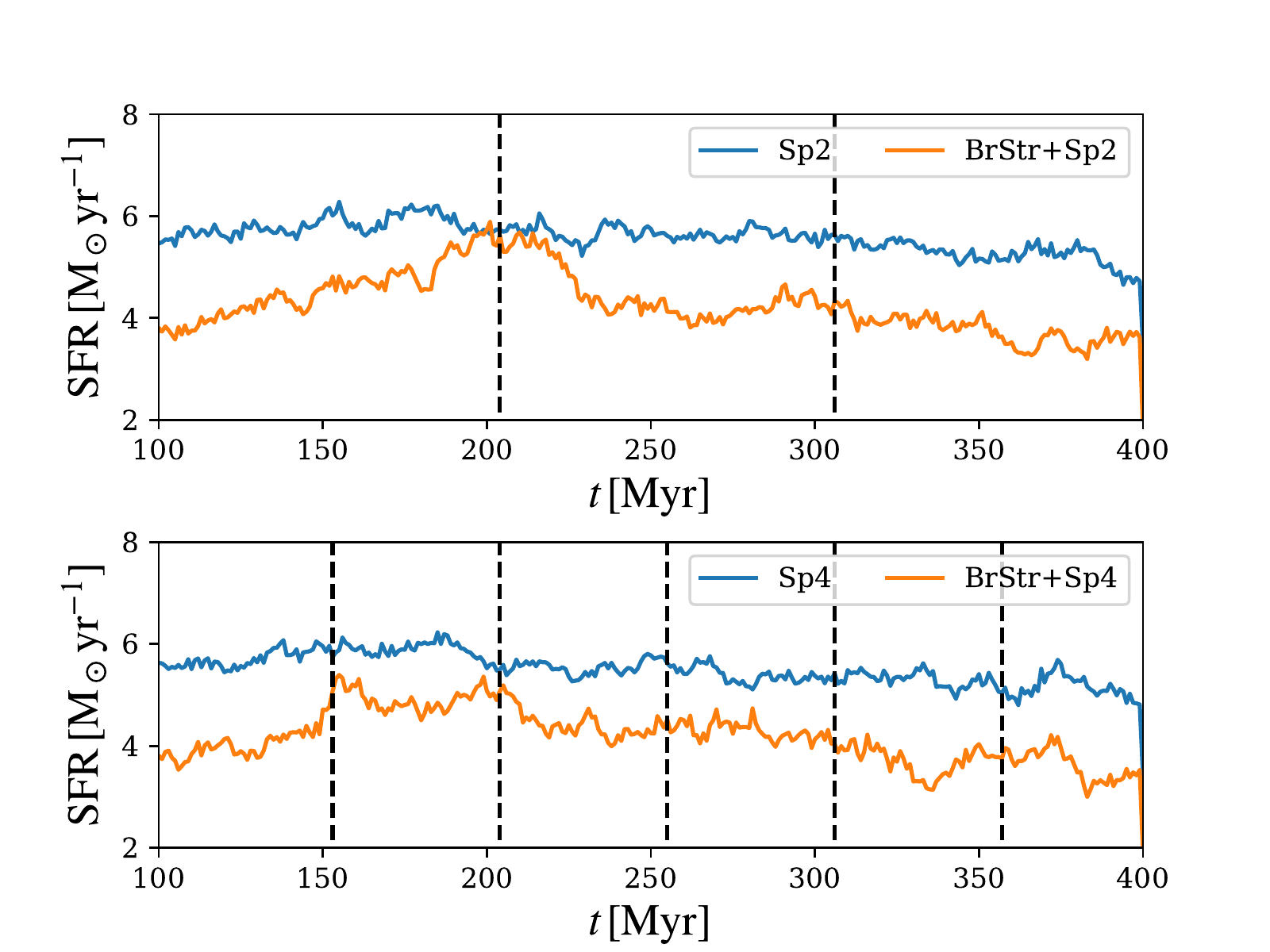}
 \caption{Star formation rate as a function of time BrStr and spiral models, with ``beat" frequencies shown as vertical dashed lines. The unbarred SFR are offset vertically by $2M_\odot {\rm yr^{-1}}$ for clarity.}
\label{StrongBarSFR}
\end{figure}

We show binned young stars as a function of azimuth in Figure\;\ref{BeatsStrong}. Very similar features are seen as in the weak bar case, with a clear time-dependence of the locations of peaks that correlate very well with the bar orientation as a function of time (e.g. compare BrStr+Sp2 at 310\,Mr when the arms and bar are coincident, and then at 370\,Myr when they have moved apart). The SFR shows a similar beat-like trend in Figure\;\ref{StrongBarSFR}, with beats separated further apart compared to Fig.\;\ref{SFRbeat} due to the slower bar pattern speed.

\bsp
\label{lastpage}
\end{document}